\documentclass[11pt]{article}
\usepackage[T1]{fontenc}
\usepackage[latin9]{inputenc}
\usepackage[a4paper]{geometry}
\usepackage[active]{srcltx}
\usepackage{textcomp}
\usepackage{amsmath}
\usepackage{amssymb}
\usepackage{esint}

\makeatletter
\pdfoutput=1 

\usepackage{jheppub}

\usepackage{etoolbox}
    
    \patchcmd{\maketitle}{\@fpheader}{}{}{}

\usepackage{amsfonts}

\usepackage{bbm}

\title{\boldmath Asymptotically flat structure of hypergravity in three spacetime dimensions}

\author[a,b]{Oscar Fuentealba,}
\author[a]{Javier Matulich,}
\author[a]{Ricardo Troncoso,}

\affiliation[a]{Centro de Estudios Cient\'{i}ficos (CECs), Av. Arturo Prat 514, Valdivia,
Chile.}
\affiliation[b]{Departamento de F\'isica, Universidad de Concepci\'on, Casilla 160-C, Concepci\'on, Chile.}

\emailAdd{fuentealba@cecs.cl}
\emailAdd{matulich@cecs.cl}
\emailAdd{troncoso@cecs.cl}

\preprint{CECS-PHY-15/03}

\abstract{The asymptotic structure of three-dimensional hypergravity without cosmological constant is analyzed. In the case of gravity minimally coupled to a spin-$5/2$ field, a consistent set of boundary conditions is proposed, being wide enough so as to accommodate a generic choice of chemical potentials associated to the global charges. The algebra of the canonical generators of the asymptotic symmetries is given by a hypersymmetric nonlinear extension of BMS$_{3}$. It is shown that the asymptotic symmetry algebra can be recovered from a subset of a suitable limit of the direct sum of the W$_{\left(2,4\right)}$ algebra with its hypersymmetric extension. The presence of hypersymmetry generators allows to construct bounds for the energy, which turn out to be nonlinear and saturate for spacetimes that admit globally-defined ``Killing vector-spinors''. The null orbifold or Minkowski spacetime can then be seen as the corresponding ground state in the case of fermions that fulfill periodic or antiperiodic boundary conditions, respectively. The hypergravity theory is also explicitly extended so as to admit parity-odd terms in the action. It is then shown that the asymptotic symmetry algebra includes an additional central charge, being proportional to the coupling of the Lorentz-Chern-Simons form. The generalization of these results in the case of gravity minimally coupled to arbitrary half-integer spin fields is also carried out. The hypersymmetry bounds are found to be given by a suitable polynomial of degree $s+\frac{1}{2}$ in the energy, where $s$ is the spin of the fermionic generators.}

\makeatother

\begin{document}
\maketitle \flushbottom

\section{Introduction}

It has been shown that the inconsistencies arising in the minimal
coupling of a massless spin-$5/2$ field to General Relativity \cite{Weinberg},
\cite{AD-ConsistencyProb}, \cite{BVHVN}, \cite{Weinberg-Witten}
can be successfully surmounted in three-dimensional spacetimes \cite{AD-3D-HYGRA}.
This theory is known as hypergravity, and it has been recently reformulated
as a Chern-Simons theory of a new extension of the Poincaré group
with fermionic generators of spin $3/2$ \cite{FMT-3D-HYGRA}. In
the case of negative cosmological constant, additional spin-4 fields
are required by consistency \cite{CLW}, \cite{Zinoviev}, \cite{Hyper-AdS},
and it can be seen that the anticommutator of the generators of the
asymptotic hypersymmetries, associated to fermionic spin-$3/2$ parameters,
leads to interesting nonlinear bounds for the bosonic global charges
of spin 2 and 4 \cite{Hyper-AdS}. The bounds saturate provided the
bosonic configurations admit globally-defined ``Killing vector-spinors''.
One of the main purposes of this paper is to show how these results
extend to the case of asymptotically flat spacetimes in hypergravity,
also in the case of arbitrary half-integer spin fields. In the next
section we briefly summarize the formulation of hypergravity as a
Chern-Simons theory for the hyper-Poincaré group in the simplest case
of fermionic spin-$5/2$ fields, while section \ref{Section3} is
devoted to explore the global hypersymmetry properties of cosmological
spacetimes and solutions with conical defects. In the case of fermions
that fulfill periodic boundary conditions, it is shown that the null
orbifold possesses a single constant Killing vector-spinor. Analogously,
for antiperiodic boundary conditions, Minkowski spacetime is singled
out as the maximally (hyper)symmetric configuration, and the explicit
expression of the globally-defined Killing vector-spinors is found.
The asymptotically flat structure of hypergravity in three spacetime
dimensions is analyzed in section \ref{Asymptotics}, where a precise
set of boundary conditions that includes ``chemical potentials'' associated
to the global charges is proposed. The algebra of the canonical generators
of the asymptotic symmetries is found to be given by a suitable hypersymmetric
nonlinear extension of the BMS$_{3}$ algebra. It is also shown that
this algebra corresponds to a subset of a suitable Inönü-Wigner contraction
of the direct sum of the W$_{\left(2,4\right)}$ algebra with its
hypersymmetric extension W$_{\left(2,\frac{5}{2},4\right)}$. The
hypersymmetry bounds that arise from the anticommutator of the fermionic
generators are found to be nonlinear, and are shown to saturate for
spacetimes that admit unbroken hypersymmetries, like the ones aforementioned.
This is explicitly carried out in section \ref{Bounds}. In section
\ref{HypergravityReloaded}, the previous analysis is performed in
the case of an extension of the hypergravity theory that includes
additional parity-odd terms in the action. It is found that the asymptotic
symmetry algebra admits an additional central charge along the Virasoro
subgroup. The results are then extended to the case of General Relativity
minimally coupled to half-integer spin fields in section \ref{HypergravityS},
including the asymptotically flat structure, and the explicit expression
of the Killing tensor-spinors. The hypersymmetry bounds are shown
to be described by a polynomial of degree $s+1/2$ in the energy,
where $s$ is the spin of the fermionic generators. We conclude in
section \ref{FinalRemarks} with some final remarks, including the
extension of these results to the case of hypergravity with additional
parity-odd terms and fermions of arbitrary half-integer spin. The
coupling of additional spin-4 fields is also addressed. Appendix \ref{A}
is devoted to our conventions, and in appendix \ref{ExactKillingV-S},
an alternative interesting form to obtain the explicit form of the
Killing vector-spinors is presented. The general form of the hyper-Poincaré
algebra is discussed in appendix \ref{hyperPoincareSpinS}, while
appendix \ref{AppendixD} includes the asymptotic hypersymmetry
algebra in the case of fermionic fields of spin $3/2$ (supergravity),
as well as for fields of spin $7/2$ and $9/2$.

\section{General Relativity minimally coupled to a spin-$5/2$ field \label{hypergravity5/2}}

It has been recently shown that the hypergravity theory of Aragone
and Deser \cite{AD-3D-HYGRA} can be reformulated as a gauge theory
of a suitable extension of the Poincaré group with fermionic spin-$3/2$
generators \cite{FMT-3D-HYGRA}. The action is described by a Chern-Simons
form, so that the dreibein, the (dualized) spin connection, and the
spin-$5/2$ field correspond to the components of a gauge field given
by
\begin{equation}
A=e^{a}P_{a}+\omega^{a}J_{a}+\psi_{a}^{\alpha}Q_{\alpha}^{a}\,,
\end{equation}
that takes values in the hyper-Poincaré algebra, being spanned by
the set $\left\{ P_{a},J_{a},Q_{\alpha}^{a}\right\} $. The fermionic
fields and generators are assumed to be $\Gamma$-traceless, i. e.,
$\Gamma^{a}\psi_{a}=0$, and $Q^{a}\Gamma_{a}=0$, so that the nonvanishing
(anti)commutation rules read
\begin{gather}
\left[J_{a},J_{b}\right]=\varepsilon_{abc}J^{c}\quad,\quad\left[J_{a},P_{b}\right]=\varepsilon_{abc}P^{c}\,,\nonumber \\
\left[J_{a},Q_{\alpha b}\right]=\frac{1}{2}\left(\Gamma_{a}\right)_{\,\,\,\,\alpha}^{\beta}Q_{\beta b}+\varepsilon_{abc}Q_{\alpha}^{c}\,,\label{eq:hyperPoincare}\\
\left\{ Q_{\alpha}^{a},Q_{\beta}^{b}\right\} =-\frac{2}{3}\left(C\Gamma^{c}\right)_{\alpha\beta}P_{c}\eta^{ab}+\frac{5}{6}\varepsilon^{abc}C_{\alpha\beta}P_{c}+\frac{1}{6}(C\Gamma^{(a})_{\alpha\beta}P^{b)}\,,\nonumber 
\end{gather}
where $C$ stands for the charge conjugation matrix. The Majorana
conjugate then reads $\bar{\psi}_{\alpha a}=\psi_{a}^{\beta}C_{\beta\alpha}$.
Since the algebra admits an invariant bilinear form, whose only nonvanishing
components are given by
\begin{equation}
\left\langle J_{a},P_{b}\right\rangle =\eta_{ab}\quad,\quad\left\langle Q_{\alpha}^{a},Q_{\beta}^{b}\right\rangle =\frac{2}{3}C_{\alpha\beta}\eta^{ab}-\frac{1}{3}\varepsilon^{abc}(C\Gamma_{c})_{\alpha\beta}\,,\label{eq:IBF}
\end{equation}
the action can be written as
\begin{equation}
I\left[A\right]=\frac{k}{4\pi}\int\left\langle AdA+\frac{2}{3}A^{3}\right\rangle \,,\label{eq:Chern-Simons}
\end{equation}
which up to a surface term reduces to
\begin{equation}
I=\frac{k}{4\pi}\int2R^{a}e_{a}+i\bar{\psi}_{a}D\psi^{a}.\label{eq:Ihyper}
\end{equation}
Here $R^{a}=d\omega^{a}+\frac{1}{2}\varepsilon^{abc}\omega_{b}\omega_{c}$
is the dual of the curvature two-form, and since the fermionic field
is $\Gamma$-traceless, its Lorentz covariant derivative fulfills
\begin{eqnarray}
D\psi^{a} & = & d\psi^{a}+\frac{1}{2}\omega^{b}\Gamma_{b}\psi^{a}+\varepsilon^{abc}\omega_{b}\psi_{c}\nonumber \\
 & = & d\psi^{a}+\frac{3}{2}\omega^{b}\Gamma_{b}\psi^{a}-\omega_{b}\Gamma^{a}\psi^{b}\,.
\end{eqnarray}

The field equations are then given by $F=dA+A^{2}=0$, whose components
read 
\begin{equation}
R^{a}=0\quad,\quad T^{a}=\frac{3}{4}i\bar{\psi}_{b}\Gamma^{a}\psi^{b}\quad,\quad D\psi^{a}=0\,,\label{eq:FE}
\end{equation}
where $T^{a}=De^{a}$ corresponds to the torsion two-form. 

Therefore, by construction, the action changes by a boundary term
under local hypersymmetry transformations spanned by $\delta A=dA+\left[A,\lambda\right]$,
with $\lambda=\epsilon_{a}^{\alpha}Q_{\alpha}^{a}$, so that the transformation
law of the fields reduces to
\begin{equation}
\delta e^{a}=\frac{3}{2}i\bar{\epsilon}_{b}\Gamma^{a}\psi^{b}\quad,\quad\delta\omega^{a}=0\quad,\quad\delta\psi^{a}=D\epsilon^{a}\,.\label{eq:TL}
\end{equation}
Note that the transformation rules of the fields in \cite{AD-3D-HYGRA}
agree with the ones in \eqref{eq:TL}, on-shell.

\section{Unbroken hypersymmetries: Killing vector-spinors \label{Section3}}

It is interesting to explore the set of bosonic solutions that possess
unbroken global hypersymmetries. According to the transformation rules
of the fields in \eqref{eq:TL}, this class of configurations has
to fulfill the following Killing vector-spinor equation:
\begin{equation}
d\epsilon^{a}+\frac{1}{2}\omega^{b}\Gamma_{b}\epsilon^{a}+\varepsilon^{abc}\omega_{b}\epsilon_{c}=0\,,\label{eq:KillingEq}
\end{equation}
where the spin-$3/2$ parameter $\epsilon^{a}$ is $\Gamma$-traceless.

As it follows from the field equations \eqref{eq:FE}, the spin connection
is locally flat, and it can then be written as $\omega=\omega^{a}J_{a}=g^{-1}dg$,
with $g=e^{\lambda^{a}J_{a}}$. Therefore, the general solution of
the Killing vector-spinor equation \eqref{eq:KillingEq} is given
by
\begin{equation}
\epsilon_{a}^{\alpha}=\left(g_{S}^{-1}\right)_{\beta}^{\alpha}\left(g_{V}\right)_{a}^{b}\eta_{b}^{\beta},\label{eq:Killing VS}
\end{equation}
where $\eta_{b}^{\beta}$ is a $\Gamma$-traceless constant vector-spinor.
Here, $g_{S}$ and $g_{V}$ stand for the same group element $g$,
but expressed in the spinor and the vector (adjoint) representations,
respectively. Since the generators of the Lorentz group in the spinor
and vector representations are given by $\left(J_{a}\right)_{\beta}^{\alpha}=\frac{1}{2}\left(\Gamma_{a}\right)_{\beta}^{\alpha}$,
and $\left(J_{a}\right)_{bc}=-\varepsilon_{abc}$, they explicitly
read 
\begin{equation}
\left(g_{S}\right)_{\beta}^{\alpha}=\exp\left[\frac{1}{2}\lambda^{a}\left(\Gamma_{a}\right)_{\beta}^{\alpha}\right]\quad,\quad\left(g_{V}\right)_{bc}=\exp\left[-\lambda^{a}\varepsilon_{abc}\right]\,.\label{eq:GroupElements}
\end{equation}

Hence, bosonic configurations that admit unbroken hypersymmetries
possess Killing vector-spinors of the form \eqref{eq:Killing VS}
provided they are globally well-defined, either for periodic or antiperiodic
boundary conditions.

\subsection{Cosmological spacetimes and solutions with conical defects \label{Cosmology}}

Let us focus on an interesting class of circularly symmetric solutions
that describe cosmological spacetimes as well as configurations with
conical defects. The latter class was introduced in \cite{DJT-3D},
\cite{Deser-Jackiw} while the former one was explored in \cite{Ezawa},
\cite{Cornalba-Costa-Cosm}, \cite{Cornalba-Costa-Orbifolds}. The
thermodynamic properties of cosmological spacetimes have been analyzed
in \cite{Barnich-Cosmology}, \cite{Bagchi.et.al-Flat Cosmology},
\cite{Gary-HSChP}, \cite{Matulich-HSChP}. As explained in \cite{Henneaux-CP3D},
\cite{BHPTT-GBH3D}, \cite{Hyper-AdS}, it is useful to express the
solution for a fixed range of the coordinates, so that the Hawking
temperature and the chemical potential for the angular momentum manifestly
appear in the metric. Hereafter we follow the conventions of \cite{Matulich-HSChP},
and for latter purposes, it is convenient to write the line element
in outgoing null coordinates, which reads
\begin{equation}
ds^{2}=-\frac{4\pi}{k}\left(\frac{\pi\mathcal{J}^{2}}{kr^{2}}-\mathcal{P}\right)\mu_{\mathcal{P}}^{2}du^{2}-2\mu_{\mathcal{P}}dudr+r^{2}\left[d\phi+\left(\mu_{\mathcal{J}}+\frac{2\pi\mu_{\mathcal{P}}\mathcal{J}}{kr^{2}}\right)du\right]^{2}\,.\label{eq:LE}
\end{equation}
Here $\mathcal{P}$ determines the mass, whose associated ``chemical
potential'' relates to the inverse Hawking temperature according to
$\mu_{\mathcal{P}}=-\beta^{-1}$. Analogously, $\mu_{\mathcal{J}}$
stands for the chemical potential associated to the angular momentum
$\mathcal{J}$. We also assume a non-diagonal form for the Minkowski
metric in a local frame, so that its nonvanishing components are given
by $\eta_{01}=\eta_{10}=\eta_{22}=1$. The dreibein can then be chosen
as
\begin{equation}
e^{0}=-dr+\frac{2\pi\mu_{\mathcal{P}}\mathcal{P}}{k}du+\frac{2\pi\mathcal{J}}{k}\left(d\phi+\mu_{\mathcal{J}}du\right)\quad,\quad e^{1}=\mu_{\mathcal{P}}du\quad,\quad e^{2}=r\left(d\phi+\mu_{\mathcal{J}}du\right)\,,
\end{equation}
and hence, the components of the dualized spin connection are given
by
\begin{equation}
\omega^{0}=\frac{2\pi\mathcal{P}}{k}\left(d\phi+\mu_{\mathcal{J}}du\right)\quad,\quad\omega^{1}=d\phi+\mu_{\mathcal{J}}du\quad,\quad\omega^{2}=0\,.\label{eq:SpinConnection}
\end{equation}
As explained at the beginning of section \ref{Section3}, since the
curvature two-form vanishes, the spin connection \eqref{eq:SpinConnection}
is locally flat, and it can then be generically written as $\omega=g^{-1}dg$,
with 
\begin{equation}
g=\exp\left[\left(J_{1}+\frac{2\pi\mathcal{P}}{k}J_{0}\right)\hat{\phi}\right]\,,\label{eq:g}
\end{equation}
and $\hat{\phi}=\phi+\mu_{\mathcal{J}}u$.

Note that in the case of $\mathcal{P}\neq0$, for the spinor and vector
representations, the group element $g$ in \eqref{eq:g} exponentiates
as

\begin{equation}
g_{S}=\cosh\left[\sqrt{\frac{\pi\mathcal{P}}{k}}\hat{\phi}\right]\mathbb{I}_{2\times2}+\sqrt{\frac{k}{\pi\mathcal{P}}}\sinh\left[\sqrt{\frac{\pi\mathcal{P}}{k}}\hat{\phi}\right]\left(J_{1}+\frac{2\pi\mathcal{P}}{k}J_{0}\right)\,,\label{eq:gsPneq0}
\end{equation}
\begin{equation}
g_{V}=\mathbb{I}_{3\times3}+\frac{1}{2}\sqrt{\frac{k}{\pi\mathcal{P}}}\sinh\left[2\sqrt{\frac{\pi\mathcal{P}}{k}}\hat{\phi}\right]\left(J_{1}+\frac{2\pi\mathcal{P}}{k}J_{0}\right)+\frac{k}{2\pi\mathcal{P}}\sinh\left[\sqrt{\frac{\pi\mathcal{P}}{k}}\hat{\phi}\right]^{2}\left(J_{1}+\frac{2\pi\mathcal{P}}{k}J_{0}\right)^{2}\,,\label{eq:gVPneq0}
\end{equation}
respectively, while for $\mathcal{P}=0$, it reduces to
\begin{equation}
g_{S}=\mathbb{I}_{2\times2}+\hat{\phi}J_{1}\quad,\quad g_{V}=\mathbb{I}_{3\times3}+\hat{\phi}J_{1}+\frac{1}{2}\hat{\phi}^{2}J_{1}^{2}\,.\label{eq:gP0}
\end{equation}

One then concludes that cosmological spacetimes, for which $\mathcal{P}>0$,
necessarily break all the hypersymmetries. Indeed, this class of solutions
cannot admit globally-defined Killing vector-spinors because, according
to \eqref{eq:gsPneq0} and \eqref{eq:gVPneq0}, the (anti)periodic
boundary conditions for the vector-spinor $\epsilon_{a}$ in \eqref{eq:Killing VS}
fail to be fulfilled. 

In the case of configurations with $\mathcal{P}=0$, equations \eqref{eq:Killing VS}
and \eqref{eq:gP0} imply that the Killing vector-spinor is constant
and satisfies:
\begin{equation}
\frac{3}{2}\Gamma_{1}\epsilon_{a}-\Gamma_{a}\epsilon_{1}=0\,,
\end{equation}
so that it fulfills periodic boundary conditions, and possesses a
single nonvanishing component given by $\epsilon_{0}^{-}=\eta_{0}^{-}$.

For the remaining case, $\mathcal{P}:=-kj^{2}/\pi<0$, describing
solutions with conical defects, the group element in both representations
reduces to
\begin{equation}
g_{S}=\cos\left[j\hat{\phi}\right]\mathbb{I}_{2\times2}+\frac{1}{j}\sin\left[j\hat{\phi}\right]\left(J_{1}-2j^{2}J_{0}\right)\,,\label{eq:gsP<0}
\end{equation}
\begin{equation}
g_{V}=\mathbb{I}_{3\times3}+\frac{1}{2j}\sin\left[2j\hat{\phi}\right]\left(J_{1}-2j^{2}J_{0}\right)+\frac{1}{2j^{2}}\sin\left[j\hat{\phi}\right]^{2}\left(J_{1}-2j^{2}J_{0}\right)^{2}\,.\label{eq:gVP<0}
\end{equation}
Therefore, this class of configurations possesses four independent
Killing vector-spinors that fulfill (anti)periodic boundary conditions
provided $j$ is a (half-)integer. The explicit form of the Killing
vector-spinors is then obtained from \eqref{eq:Killing VS}, where
$g_{S}$ and $g_{V}$ are given by eqs. \eqref{eq:gsP<0} and \eqref{eq:gVP<0}.
Note that this is the maximum number of hypersymmetries. Indeed, for
these configurations the holonomy of the spin connection becomes trivial,
which in the spinor representation means that $g_{S}^{-1}(\hat{\phi})g_{S}(\hat{\phi}+2\pi)=-\mathbb{I}_{2\times2}$,
while in the vector representation the condition reads $g_{V}^{-1}(\hat{\phi})g_{V}(\hat{\phi}+2\pi)=\mathbb{I}_{3\times3}$.
It is worth pointing out that if $j$ were different from a (half-)integer,
the configurations would not solve the field equations in vacuum.
This is because they would possess a conical singularity at the origin,
and hence they should necessarily be supported by an external source.

As it occurs in the case of supersymmetry, it is natural to expect
that the bosonic global charges fulfill suitable bounds that turn
out to be saturated for configurations that possess unbroken hypersymmetries.
Indeed, as shown in \cite{BDMT}, the bounds that correspond to three-dimensional
supergravity with asymptotically flat boundary conditions certainly
do so. Actually, the bounds also exclude conical surplus solutions,
in particular those whose angular coordinate ranges from zero to $4\pi j$,
with $j>1/2$, despite they are maximally supersymmetric. When a negative
cosmological constant is considered, this is also the case not only
for supergravity \cite{Coussaert-Henneaux}, but also for hypergravity
\cite{Hyper-AdS}, where in the latter case the bounds turn out to
be nonlinear. Thus, one of the main purposes of the following sections
is showing how these results can be extended to the case of hypergravity
endowed with a suitable set of asymptotically flat boundary conditions,
as well as how to recover them in the vanishing cosmological constant
limit.

\section{Asymptotically flat behaviour and the hyper-BMS$_{3}$ algebra \label{Asymptotics}}

Let us introduce a suitable set of asymptotic conditions that allows
to describe the dynamics of asymptotically flat spacetimes in hypergravity.
The set must be relaxed enough so as to accommodate the solutions
of interest that have been described in section \ref{Cosmology},
and simultaneously, restricted in an appropriate way in order to ensure
finiteness of the canonical generators associated to the asymptotic
symmetries. In the case of pure General Relativity, a consistent set
of boundary conditions indeed exists, whose asymptotic symmetry algebra
corresponds to BMS$_{3}$ with a nontrivial central extension \cite{Ashtekar-BMS},
\cite{Barnich-Compere}, \cite{Barnich-Cedric}. These results have
been extended to the case of supergravity \cite{BDMT}, as well as
for General Relativity coupled to higher spin fields \cite{ABFGR-HSF3D},
\cite{GMPT-HSF3D}, \cite{Gary-HSChP}, \cite{Matulich-HSChP}. In
order to carry out this task in hypergravity, we take advantage of
the Chern-Simons formulation of the theory, depicted in section \ref{hypergravity5/2}.
Since the hypersymmetry generators are $\Gamma-$traceless, it is
useful to get rid of $Q_{2}=Q_{1}\Gamma_{0}-Q_{0}\Gamma_{1}$, so
that once the remaining generators are relabeled according to
\begin{gather}
\hat{\mathcal{J}}_{-1}=-2J_{0}\quad,\quad\hat{\mathcal{J}}_{1}=J_{1}\quad,\quad\hat{\mathcal{J}}_{0}=J_{2}\,,\nonumber \\
\hat{\mathcal{P}}_{-1}=-2P_{0}\quad,\quad\hat{\mathcal{P}}_{1}=P_{1}\quad,\quad\hat{\mathcal{P}}_{0}=P_{2}\,,\nonumber \\
\hat{\mathcal{Q}}_{-\frac{3}{2}}=2^{\frac{5}{4}}\sqrt{3}Q_{+0}\quad,\quad\hat{\mathcal{Q}}_{-\frac{1}{2}}=2^{\frac{3}{4}}\sqrt{3}Q_{-0}\,,\\
\hat{\mathcal{Q}}_{\frac{1}{2}}=-2^{\frac{1}{4}}\sqrt{3}Q_{+1}\quad,\quad\hat{\mathcal{Q}}_{\frac{3}{2}}=-2^{-\frac{1}{4}}\sqrt{3}Q_{-1}\,,\nonumber 
\end{gather}
the hyper-Poincaré algebra \eqref{eq:hyperPoincare} reads
\begin{eqnarray}
\left[\hat{\mathcal{J}}_{m},\hat{\mathcal{J}}_{n}\right] & = & \left(m-n\right)\hat{\mathcal{J}}_{m+n}\,,\nonumber \\
\left[\hat{\mathcal{J}}_{m},\hat{\mathcal{P}}_{n}\right] & = & \left(m-n\right)\hat{\mathcal{P}}_{m+n}\,,\nonumber \\
\left[\hat{\mathcal{J}}_{m},\hat{\mathcal{Q}}_{p}\right] & = & \left(\frac{3m}{2}-p\right)\hat{\mathcal{Q}}_{m+p}\,,\label{eq:hyperPoincareModos}\\
\left\{ \hat{\mathcal{Q}}_{p},\hat{\mathcal{Q}}_{q}\right\}  & = & \frac{1}{4}\left(6p^{2}-8pq+6q^{2}-9\right)\hat{\mathcal{P}}_{p+q}\,,\nonumber 
\end{eqnarray}
with $m,n=\pm1,0$, and $p,q=\pm\frac{1}{2},\pm\frac{3}{2}$.

Thus, following the lines of \cite{CH-vanDriel}, and as explained
in \cite{BDMT}, \cite{GMPT-HSF3D}, the radial dependence of the
asymptotic form of the gauge field can be gauged away by a suitable
group element of the form $h=e^{\frac{r}{2}\hat{\mathcal{P}}_{-1}}$,
so that
\begin{equation}
A=h^{-1}ah+h^{-1}dh\,,\label{eq:Aconh}
\end{equation}
and hence, the remaining analysis can be entirely performed in terms
of the connection $a=a_{u}du+a_{\phi}d\phi$, that depends only on
time and the angular coordinate. As explained in \cite{Henneaux-CP3D},
\cite{BHPTT-GBH3D}, one starts prescribing the asymptotic form of
the dynamical gauge field at a fixed time slice with $u=u_{0}$, so
that the asymptotic fall-off of $a_{\phi}$ is assumed to be such
that the deviations with respect to the reference background go along
the highest weight generators of \eqref{eq:hyperPoincareModos}. Choosing
the reference background to be given by the null orbifold \cite{Horowitz-Steif},
that corresponds to the configuration in \eqref{eq:LE} with $\mathcal{J}=P=0$,
the asymptotic form of the dynamical field reads
\begin{equation}
a_{\phi}=\hat{\mathcal{J}}_{1}-\frac{\pi}{k}\left(\mathcal{J}\hat{\mathcal{P}}_{-1}+\mathcal{P}\hat{\mathcal{J}}_{-1}-\frac{\psi}{3}\hat{\mathcal{Q}}_{-\frac{3}{2}}\right)\,,\label{eq:aphi}
\end{equation}
where $\mathcal{J}$, $\mathcal{P}$ and $\psi$ stand for arbitrary
functions of $u$, $\phi$. The asymptotic symmetries then correspond
to gauge transformations $\delta a=d\lambda+\left[a,\lambda\right]$
that preserve the form of \eqref{eq:aphi}. Therefore, the hyper-Poincaré-valued
parameter $\lambda$ is found to depend on three arbitrary functions
of $u$ and $\phi$, so that
\begin{equation}
\lambda=T\hat{\mathcal{P}}_{1}+Y\hat{\mathcal{J}}_{1}+\mathcal{E}\hat{\mathcal{Q}}_{\frac{3}{2}}+\eta_{\left(\frac{3}{2}\right)}\left[T,Y,\mathcal{E}\right]\,,\label{eq:lambda}
\end{equation}
where $\mathcal{E}$ is Grassmann-valued, and 
\begin{eqnarray}
\eta_{\left(\frac{3}{2}\right)}\left[T,Y,\mathcal{E}\right] & = & -T\mbox{\ensuremath{'}}\hat{\mathcal{P}}_{0}-Y\mbox{\ensuremath{'}}\hat{\mathcal{J}}_{0}-\mathcal{E}\mbox{\ensuremath{'}}\hat{\mathcal{Q}}_{\frac{1}{2}}-\frac{1}{2}\left(\frac{2\pi}{k}Y\mathcal{P}-Y\mbox{\ensuremath{''}}\right)\hat{\mathcal{J}}_{-1}\nonumber \\
 &  & -\frac{\pi}{k}\left(T\mathcal{P}+Y\mathcal{J}-\frac{3}{2}i\psi\mathcal{E}-\frac{k}{2\pi}T\mbox{\ensuremath{''}}\right)\hat{\mathcal{P}}_{-1}-\frac{1}{2}\left(\frac{3\pi}{k}\mathcal{E}\mathcal{P}-\mathcal{E}\mbox{\ensuremath{''}}\right)\hat{\mathcal{Q}}_{-\frac{1}{2}}\nonumber \\
 &  & -\frac{\pi}{3k}\left(Y\psi-\frac{7}{2}\mathcal{E}\mbox{\ensuremath{'}}\mathcal{P}-\frac{3}{2}\mathcal{E}\mathcal{P}\mbox{\ensuremath{'}}+\frac{k}{2\pi}\mathcal{E}\mbox{\ensuremath{'''}}\right)\hat{\mathcal{Q}}_{-\frac{3}{2}}\,;\label{eq:eta5/2}
\end{eqnarray}
while the transformation law of the fields reads
\begin{eqnarray}
\delta\mathcal{P} & = & 2\mathcal{P}Y\mbox{\ensuremath{'}}+\mbox{\ensuremath{\mathcal{P}}}\mbox{\ensuremath{'}}Y-\frac{k}{2\pi}Y\mbox{\ensuremath{'''}}\,,\nonumber \\
\delta\mathcal{\mathcal{J}} & = & 2\mathcal{J}Y\mbox{\ensuremath{'}}+\mathcal{J}\mbox{\ensuremath{'}}Y+2\mathcal{P}T\mbox{\ensuremath{'}}+\mathcal{P}\mbox{\ensuremath{'}}T-\frac{k}{2\pi}T\mbox{\ensuremath{'''}}+\frac{5}{2}i\psi\mathcal{E}\mbox{\ensuremath{'}}+\frac{3}{2}i\psi\mbox{\ensuremath{'}}\mathcal{E}\,,\label{eq:dfields}\\
\delta\psi & = & \frac{5}{2}\psi Y\mbox{\ensuremath{'}}+\psi\mbox{\ensuremath{'}}Y-\frac{9\pi}{2k}\mathcal{P}^{2}\mathcal{E}+\frac{3}{2}\mathcal{P}\mbox{\ensuremath{''}}\mathcal{E}+5\mathcal{P}\mbox{\ensuremath{'}}\mathcal{E}\mbox{\ensuremath{'}}+5\mathcal{P}\mathcal{E}\mbox{\ensuremath{''}}-\frac{k}{2\pi}\mathcal{E}\mbox{\ensuremath{''''}}\,.\nonumber 
\end{eqnarray}
Hereafter, prime stands for $\partial_{\phi}$. Since the time evolution
of $a_{\phi}$ corresponds to a gauge transformation parametrized
by the Lagrange multiplier $a_{u}$, its asymptotic form will be maintained
along different time slices provided $a_{u}$ is of the allowed form,
i. e.,
\begin{equation}
a_{u}=\lambda\left[\mu_{\mathcal{P}},\mu_{\mathcal{J}},\mu_{\mathcal{\psi}}\right]\,,\label{eq:au}
\end{equation}
where the chemical potentials $\mu_{\mathcal{P}},\mu_{\mathcal{J}},\mu_{\mathcal{\psi}}$
stand for arbitrary functions of $u$, $\phi$, that are assumed to
be fixed at the boundary. Consistency then demands that the field
equations, which now reduce to
\begin{eqnarray}
\dot{\mathcal{P}} & = & 2\mathcal{P}\mu_{\mathcal{J}}\mbox{\ensuremath{'}}+\mbox{\ensuremath{\mathcal{P}}}\mbox{\ensuremath{'}}\mu_{\mathcal{J}}-\frac{k}{2\pi}\mu_{\mathcal{J}}\mbox{\ensuremath{'''}}\,,\nonumber \\
\dot{\mathcal{\mathcal{J}}} & = & 2\mathcal{J}\mu_{\mathcal{J}}\mbox{\ensuremath{'}}+\mathcal{J}\mbox{\ensuremath{'}}\mu_{\mathcal{J}}+2\mathcal{P}\mu_{\mathcal{P}}\mbox{\ensuremath{'}}+\mathcal{P}\mbox{\ensuremath{'}}\mu_{\mathcal{P}}-\frac{k}{2\pi}\mu_{\mathcal{P}}\mbox{\ensuremath{'''}}+\frac{5}{2}i\psi\mu_{\mathcal{\psi}}\mbox{\ensuremath{'}}+\frac{3}{2}i\psi\mbox{\ensuremath{'}}\mu_{\mathcal{\psi}}\,,\\
\dot{\psi} & = & \frac{5}{2}\psi\mu_{\mathcal{J}}\mbox{\ensuremath{'}}+\psi\mbox{\ensuremath{'}}\mu_{\mathcal{J}}-\frac{9\pi}{2k}\mathcal{P}^{2}\mu_{\mathcal{\psi}}+\frac{3}{2}\mathcal{P}\mbox{\ensuremath{''}}\mu_{\mathcal{\psi}}+5\mathcal{P}\mbox{\ensuremath{'}}\mu_{\mathcal{\psi}}\mbox{\ensuremath{'}}+5\mathcal{P}\mu_{\mathcal{\psi}}\mbox{\ensuremath{''}}-\frac{k}{2\pi}\mu_{\mathcal{\psi}}\mbox{\ensuremath{''''}}\,,\nonumber 
\end{eqnarray}
have to hold in the asymptotic region, while the parameters of the
asymptotic symmetries fulfill the following conditions
\begin{eqnarray}
\dot{Y} & = & \mu_{\mathcal{J}}Y\mbox{\ensuremath{'}}-\mu_{\mathcal{J}}\mbox{\ensuremath{'}}Y\,,\nonumber \\
\dot{T} & = & \mu_{\mathcal{J}}T\mbox{\ensuremath{'}}-\mu_{\mathcal{J}}\mbox{\ensuremath{'}}T+\mu_{\mathcal{P}}Y\mbox{\ensuremath{'}}-\mu_{\mathcal{P}}\mbox{\ensuremath{'}}Y+\frac{9\pi}{k}i\mu_{\psi}\mathcal{E}\mathcal{P}-\frac{3}{2}i\mu_{\psi}\mbox{\ensuremath{''}}\mathcal{E}+2i\mu_{\psi}\mbox{\ensuremath{'}}\mathcal{E}\mbox{\ensuremath{'}}-\frac{3}{2}i\mu_{\psi}\mathcal{E}\mbox{\ensuremath{''}}\,,\label{eq:ChiralityConditions}\\
\dot{\mathcal{E}} & = & \frac{3}{2}\mu_{\psi}Y\mbox{\ensuremath{'}}-\mu_{\psi}\mbox{\ensuremath{'}}Y-\frac{3}{2}\mu_{\mathcal{J}}\mbox{\ensuremath{'}}\mathcal{E}+\mu_{\mathcal{J}}\mathcal{E}\mbox{\ensuremath{'}}\,,\nonumber 
\end{eqnarray}
which are needed in order to ensure that the global charges are conserved.%
\footnote{Since global symmetries are necessarily contained within the asymptotic
ones, these results provide an interesting alternative path to find
the explicit expression of the Killing vector-spinors. See appendix
\ref{ExactKillingV-S}.%
}

Following the Regge-Teitelboim approach \cite{Regge-Teitelboim},
the variation of the canonical generators is found to be generically
given by 
\begin{equation}
\delta Q\left[\lambda\right]=-\frac{k}{2\pi}\int\left\langle \lambda\delta a_{\phi}\right\rangle d\phi\,,\label{eq:CS-Charge}
\end{equation}
which by virtue of \eqref{eq:aphi} and \eqref{eq:lambda}, up to
an arbitrary constant without variation, integrate as
\begin{equation}
Q\left[T,Y,\mathcal{E}\right]=-\int\left(T\mathcal{P}+Y\mathcal{J}-i\mathcal{E}\psi\right)d\phi\,.\label{eq:Charge5/2}
\end{equation}
It is worth highlighting that the global charges are manifestly independent
of the radial coordinate $r$. Therefore, the boundary can be located
at an arbitrary fixed value $r=r_{0}$, and it corresponds to a timelike
surface with the topology of a cylinder. 

Since the Poisson brackets fulfill $\left\{ Q\left[\lambda_{1}\right],Q\left[\lambda_{2}\right]\right\} =\delta_{\lambda_{2}}Q\left[\lambda_{1}\right]$,
the algebra of the canonical generators can be directly obtained from
the transformation law of the fields in \eqref{eq:dfields}. Expanding
in Fourier modes, $X=\frac{1}{2\pi}\sum_{n}X_{n}e^{in\phi}$, the
nonvanishing Poisson brackets read

\begin{align}
i\left\{ \mathcal{J}_{m},\mathcal{J}_{n}\right\}  & =\left(m-n\right)\mathcal{J}_{m+n}\,,\nonumber \\
i\left\{ \mathcal{J}_{m},\mathcal{P}_{n}\right\}  & =\left(m-n\right)\mathcal{P}_{m+n}+km^{3}\delta_{m+n,0}\,,\nonumber \\
i\left\{ \mathcal{J}_{m},\psi_{n}\right\}  & =\left(\frac{3m}{2}-n\right)\psi_{m+n}\,,\label{eq:hyperBMS}\\
i\left\{ \psi_{m},\psi_{n}\right\}  & =\frac{1}{2}\left(3m^{2}-4mn+3n^{2}\right)\mathcal{P}_{m+n}+\frac{9}{4k}\sum_{q}\mathcal{P}_{m+n-q}\mathcal{P}_{q}+km^{4}\delta_{m+n,0}\,,\nonumber 
\end{align}
where the modes of the generators $\psi_{m}$ are labeled by (half-)integers
when the fermions fulfill (anti)periodic boundary conditions. 

It is then clear that, with respect to $\mathcal{J}_{m}$, the conformal
weight of the generators $\mathcal{P}_{m}$ and $\psi_{n}$, is given
by 2 and $5/2$, respectively. Note that the subset spanned by $\mathcal{J}_{m}$
and $\mathcal{P}_{m}$ corresponds to the BMS$_{3}$ algebra of General
Relativity with the same central extension, and hence \eqref{eq:hyperBMS}
stands for its hypersymmetric extension that is manifestly nonlinear. 

It is useful to perform the following shift in the generators: 
\begin{equation}
\mathcal{P}_{n}\rightarrow\mathcal{P}_{n}-\frac{k}{2}\delta_{n,0}\,,\label{eq:Shift}
\end{equation}
so that the algebra now reads
\begin{eqnarray}
i\left\{ \mathcal{J}_{m},\mathcal{J}_{n}\right\}  & = & \left(m-n\right)\mathcal{J}_{m+n}\,,\nonumber \\
i\left\{ \mathcal{J}_{m},\mathcal{P}_{n}\right\}  & = & \left(m-n\right)\mathcal{P}_{m+n}+km\left(m^{2}-1\right)\delta_{m+n,0}\,,\nonumber \\
i\left\{ \mathcal{J}_{m},\mathcal{\psi}_{n}\right\}  & = & \left(\frac{3m}{2}-n\right)\psi_{m+n}\,,\label{eq:hyperBMS2}\\
i\left\{ \mathcal{\mathcal{\psi}}_{m},\mathcal{\psi}_{n}\right\}  & = & \frac{1}{4}\left(6m^{2}-8mn+6n^{2}-9\right)\mathcal{P}_{m+n}+\frac{9}{4k}\sum_{q}\mathcal{P}_{m+n-q}\mathcal{P}_{q}\nonumber \\
 &  & +k\left(m^{2}-\frac{9}{4}\right)\left(m^{2}-\frac{1}{4}\right)\delta_{m+n,0}\,,\nonumber 
\end{eqnarray}
in agreement with the result that has been recently anticipated in
\cite{FMT-3D-HYGRA}. Indeed, dropping the nonlinear terms in \eqref{eq:hyperBMS2},
when the fermions fulfill antiperiodic boundary conditions, the wedge
algebra, which is spanned by the subset of $\left\{ \mathcal{J}_{m},\mathcal{P}_{m},\psi_{n}\right\} $
with $m=\pm1,0$ and $n=\pm3/2,\pm1/2$, reduces to the hyper-Poincaré
algebra in eq. \eqref{eq:hyperPoincareModos}.

It can also be seen that the hyper-BMS$_{3}$ algebra \eqref{eq:hyperBMS}
turns out to be a subset of a precise Inönü-Wigner contraction of
the direct sum of the W$_{\left(2,4\right)}$ algebra with its hypersymmetric
extension W$_{\left(2,\frac{5}{2},4\right)}$. This is the main subject
of the next subsection.

\subsection{Flat limit of the asymptotic symmetry algebra from the case of negative
cosmological constant \label{FlatLimit}}

It has been recently shown that the asymptotic symmetries of three-dimensional
hypergravity with negative cosmological constant are spanned by two
copies of the classical limit of the WB$_{2}$ algebra \cite{Hyper-AdS}.
This algebra is also known as W$_{\left(2,\frac{5}{2},4\right)}$
and corresponds to the hypersymmetric extension of W$_{\left(2,4\right)}$
\cite{FST-WB2}, \cite{BKS-W2}. The hypergravity theory that was
discussed in \cite{Hyper-AdS} possesses the minimum number of hypersymmetries
in each sector, so that the gauge group is given by $OSp\left(1|4\right)\otimes OSp\left(1|4\right)$.
In analogy with the case of three-dimensional supergravity \cite{Achucarro:1987vz},
one may say that the theory aforementioned corresponds to the $\mathcal{N}=\left(1,1\right)$
AdS$_{3}$ hypergravity. In this sense, there are two inequivalent
minimal locally hypersymmetric extensions of General Relativity with
negative cosmological constant, which correspond to the $\left(1,0\right)$
and the $\left(0,1\right)$ theories. It is then simple to verify
that both minimal theories possess the same vanishing cosmological
constant limit, and hence in order to proceed with the analysis we
will consider the $\left(0,1\right)$ one, whose gauge group is given
by $Sp\left(4\right)\otimes OSp\left(1|4\right)$. According to \cite{Hyper-AdS},
the asymptotic symmetry algebra of the minimal hypergravity theory
with negative cosmological constant then corresponds to W$_{\left(2,4\right)}$$\oplus$W$_{\left(2,\frac{5}{2},4\right)}$. 

The classical limit of the W$_{\left(2,\frac{5}{2},4\right)}$ algebra
reads
\begin{eqnarray}
i\left\{ \mathcal{L}_{m},\mathcal{L}_{n}\right\}  & = & \left(m-n\right)\mathcal{L}_{m+n}+\frac{\kappa}{2}m^{3}\delta_{m+n,0}\,,\nonumber \\
i\left\{ \mathcal{L}_{m},\mathcal{U}_{n}\right\}  & = & \left(3m-n\right)\mathcal{U}_{m+n}\,,\nonumber \\
i\left\{ \mathcal{L}_{m},\Psi_{n}\right\}  & = & \left(\frac{3m}{2}-n\right)\Psi_{m+n}\,,\nonumber \\
i\left\{ \mathcal{U}_{m},\mathcal{U}_{n}\right\}  & = & \frac{1}{2^{2}3^{2}}\left(m-n\right)\left(3m^{4}-2m^{3}n+4m^{2}n^{2}-2mn^{3}+3n^{4}\right)\mathcal{L}_{m+n}\nonumber \\
 &  & +\frac{1}{6}\left(m-n\right)\left(m^{2}-mn+n^{2}\right)\mathcal{U}_{m+n}-\frac{2^{3}3\pi}{\kappa}\left(m-n\right)\Lambda_{m+n}^{\left(6\right)}\label{eq:WB2}\\
 &  & -\frac{7^{2}\pi}{3^{2}\kappa}\left(m-n\right)\left(m^{2}+4mn+n^{2}\right)\Lambda_{m+n}^{\left(4\right)}+\frac{\kappa}{2^{3}3^{2}}m^{7}\delta_{m+n,0}\,,\nonumber \\
i\left\{ \mathcal{U}_{m},\Psi_{n}\right\}  & = & \frac{1}{2^{2}3}\left(m^{3}-4m^{2}n+10mn^{2}-20n^{3}\right)\Psi_{m+n}-\frac{23\pi}{3\kappa}i\Lambda_{m+n}^{\left(11/2\right)}\nonumber \\
 &  & +\frac{\pi}{3\kappa}\left(23m-82n\right)\Lambda_{m+n}^{\left(9/2\right)}\,,\nonumber \\
i\left\{ \Psi_{m},\Psi_{n}\right\}  & = & \mathcal{U}_{m+n}+\frac{1}{2}\left(m^{2}-\frac{4}{3}mn+n^{2}\right)\mathcal{L}_{m+n}+\frac{3\pi}{\kappa}\Lambda_{m+n}^{\left(4\right)}+\frac{\kappa}{6}m^{4}\delta_{m+n,0}\,,\nonumber 
\end{eqnarray}
where the fermionic modes are labeled by (half-)integers in the case
of (anti)periodic boundary conditions, and $\Lambda_{m}^{\left(l\right)}=\int\Lambda^{\left(l\right)}e^{-im\phi}d\phi$
stand for the mode expansion of the nonlinear terms, given by 
\begin{eqnarray}
\Lambda^{\left(4\right)} & = & \mathcal{L}^{2}\,,\\
\Lambda^{\left(9/2\right)} & = & \mathcal{L}\Psi\,,\\
\Lambda^{\left(11/2\right)} & = & \frac{27}{23}\mathcal{L}\mbox{\ensuremath{'}}\Psi\,,\\
\Lambda^{\left(6\right)} & = & -\frac{7}{18}\mathcal{U}\mathcal{L}-\frac{8\pi}{3\kappa}\mathcal{L}^{3}+\frac{295}{432}\left(\mathcal{L}\mbox{\ensuremath{'}}\right)^{2}+\frac{22}{27}\mathcal{L}\mbox{\ensuremath{''}}\mathcal{L}+\frac{25}{12}i\Psi\Psi\mbox{\ensuremath{'}}\,.
\end{eqnarray}
The bosonic generators $\mathcal{L}_{m}$ and $\mathcal{U}_{m}$ span
the W$_{\left(2,4\right)}$ subalgebra. 

In order to take the vanishing cosmological constant limit of the
asymptotic symmetry algebra of the minimal theory, given by W$_{\left(2,4\right)}$$\oplus$W$_{\left(2,\frac{5}{2},4\right)}$,
it is useful to perform the following change of basis:
\[
\mathcal{P}_{n}=\frac{1}{\ell}\left(\mathcal{L}_{n}^{+}+\mathcal{L}_{-n}^{-}\right)\quad,\quad\mathcal{J}_{n}=\mathcal{L}_{n}^{+}-\mathcal{L}_{-n}^{-}\,,
\]
\begin{equation}
\mathcal{W}_{n}=\frac{1}{\sqrt{\ell}}\left(\mathcal{U}_{n}^{+}+\mathcal{U}_{-n}^{-}\right)\quad,\quad\mathcal{V}_{n}=\frac{1}{\sqrt{\ell}}\left(\mathcal{U}_{n}^{+}-\mathcal{U}_{-n}^{-}\right)\quad,\quad\psi_{n}=\sqrt{\frac{6}{\ell}}\Psi_{n}^{+}\,,\label{eq:Contraction1}
\end{equation}
where $\mathcal{L}_{n}^{-}$, $\mathcal{U}_{n}^{-}$ stand for the
generators of the (left) W$_{\left(2,4\right)}$ algebra, and $\mathcal{L}_{n}^{+}$,
$\mathcal{U}_{n}^{+}$, $\psi_{n}^{+}$ span the (right) W$_{\left(2,\frac{5}{2},4\right)}$
algebra. Therefore, rescaling the level according to $\kappa=k\ell$,
in the large AdS radius limit, $\ell\rightarrow\infty$, one obtains
that the nonvanishing brackets of the contracted algebra read 
\begin{eqnarray}
i\left\{ \mathcal{J}_{m},\mathcal{J}_{n}\right\}  & = & \left(m-n\right)\mathcal{J}_{m+n}\,,\nonumber \\
i\left\{ \mathcal{J}_{m},\mathcal{P}_{n}\right\}  & = & \left(m-n\right)\mathcal{P}_{m+n}+km^{3}\delta_{m+n,0}\,,\nonumber \\
i\left\{ \mathcal{J}_{m},\mathcal{W}_{n}\right\}  & = & \left(3m-n\right)\mathcal{W}_{m+n}\,,\nonumber \\
i\left\{ \mathcal{J}_{m},\mathcal{V}_{n}\right\}  & = & \left(3m-n\right)\mathcal{V}_{m+n}\,,\nonumber \\
i\left\{ \mathcal{V}_{m},\mathcal{W}_{n}\right\}  & = & \frac{1}{2^{2}3^{2}}\left(m-n\right)\left(3m^{4}-2m^{3}n+4m^{2}n^{2}-2mn^{3}+3n^{4}\right)\mathcal{P}_{m+n}\label{eq:InonuHyperBMS3}\\
 &  & -\frac{2^{3}\pi}{k}\left(m-n\right)\tilde{\Lambda}_{m+n}^{\left(6\right)}-\frac{7^{2}}{3^{2}4k}\left(m-n\right)\left(m^{2}+4mn+n^{2}\right)\sum_{q}\mathcal{P}_{m+n-q}\mathcal{P}_{q}\nonumber \\
 &  & +\frac{k}{2^{2}3^{2}}m^{7}\delta_{m+n,0}\,,\nonumber \\
i\left\{ \mathcal{J}_{m},\mathcal{\psi}_{n}\right\}  & = & \left(\frac{3m}{2}-n\right)\psi_{m+n}\,,\nonumber \\
i\left\{ \psi_{m},\psi_{n}\right\}  & = & \frac{1}{2}\left(3m^{2}-4mn+3n^{2}\right)\mathcal{P}_{m+n}+\frac{9}{4k}\sum_{q}\mathcal{P}_{m+n-q}\mathcal{P}_{q}+km^{4}\delta_{m+n,0}\,,\nonumber 
\end{eqnarray}
with
\begin{equation}
\tilde{\Lambda}^{\left(6\right)}=-\frac{7}{12}\mathcal{W}\mathcal{P}-\frac{2\pi}{k}\mathcal{P}^{3}+\frac{295}{288}\left(\mathcal{P}\mbox{\ensuremath{'}}\right)^{2}+\frac{11}{9}\mathcal{P}\mathcal{P}\mbox{\ensuremath{''}}\,.
\end{equation}

It is then apparent that one can consistently get rid of the (conformal)
spin-4 generators $\mathcal{V}_{m}$, $\mathcal{W}_{n}$, since the
Inönü-Wigner contraction of W$_{\left(2,4\right)}$$\oplus$W$_{\left(2,\frac{5}{2},4\right)}$
in eq. \eqref{eq:InonuHyperBMS3} possesses a subset spanned by $\left\{ \mathcal{P}_{m},\mathcal{J}_{m},\psi_{m}\right\} $,
which precisely corresponds to the hyper-BMS$_{3}$ algebra in \eqref{eq:hyperBMS}.
Note that this is just a reflection of the fact that in the vanishing
cosmological constant limit, the hypergravity theory can be consistently
formulated without the need of spin-4 fields.

\section{Hypersymmetry bounds \label{Bounds}}

In the case of hypergravity with negative cosmological constant, it
has been recently shown that the anticommutator of the generators
of the asymptotic hypersymmetries implies the existence of interesting
nonlinear bounds for the bosonic charges, that saturate for configurations
that admit unbroken hypersymmetries \cite{Hyper-AdS}. In this section,
following these lines, we explicitly show that this is also the case
for hypergravity with asymptotically flat boundary conditions. In
order to perform this task, it is useful to assume that the bosonic
global charges are just determined by the zero modes. Indeed, as explained
in \cite{BHPTT-GBH3D}, a generic bosonic configuration can be brought
to the \textquotedblleft rest frame\textquotedblright{} through the
action of suitable elements of the asymptotic symmetry algebra. The
searched for bounds can then be found along the same semi-classical
reasoning as in the case of supergravity \cite{Deser-Teitelboim},
\cite{Teitelboim_Susy}, \cite{Witten-positivity}, \cite{Abott-Deser},
\cite{Hull:1983ap}, \cite{Teitelboim:1984kf}, \cite{Coussaert-Henneaux}.
Hence, the fermionic bracket in \eqref{eq:hyperBMS} becomes an anticommutator,
which in the rest frame, and for $m=-n=p$, reads
\begin{equation}
\frac{1}{2\pi}\left(\hat{\mathcal{\psi}}_{p}\hat{\mathcal{\psi}}_{-p}+\hat{\psi}_{-p}\hat{\psi}_{p}\right)=5p^{2}\hat{\mathcal{P}}+\frac{9\pi}{2k}\hat{\mathcal{P}}^{2}+\frac{k}{2\pi}p^{4}\geq0\,,\label{eq:QQhat}
\end{equation}
with $\hat{\mathcal{P}}_{0}=2\pi\hat{\mathcal{P}}$. Thus, since the
left-hand side of \eqref{eq:QQhat} is a positive-definite hermitian
operator, in the classical limit, and for any value of the (half-)integer
$p$, the energy has to fulfill the following bounds: 
\begin{equation}
\left(p^{2}+\frac{9\pi}{k}\mathcal{P}\right)\left(p^{2}+\frac{\pi}{k}\mathcal{P}\right)\geq0\,,\label{eq:Bound}
\end{equation}
which are manifestly nonlinear.

Note that for any configuration with $\mathcal{P}>0$, the bounds
in \eqref{eq:Bound} are automatically fulfilled, but never saturate.
Indeed, this is the case of the cosmological spacetimes in \eqref{eq:LE},
which goes by hand with the fact that they do not admit globally-defined
Killing vector-spinors, and hence, break all the hypersymmetries.

These bounds are also clearly fulfilled in the case of $\mathcal{P}=0$,
and for fermions with periodic boundary conditions, the one for $p=0$
is saturated. This relates to the fact that this class of configurations,
that includes the null orbifold, possesses a single unbroken hypersymmetry
spanned by a constant Killing vector-spinor.

In the case of $\mathcal{P}<0$, the class of smooth configurations
are the ones for which the holonomy of the connection around an angular
cycle is trivial. This means that they are maximally (hyper)symmetric,
and then possess four Killing vector-spinors. As explained in section
\ref{Cosmology}, their energy is given by $\mathcal{P}=-kj^{2}/\pi$,
and the bounds in \eqref{eq:Bound} then reduce to
\begin{equation}
\left(p^{2}-9j^{2}\right)\left(p^{2}-j^{2}\right)\geq0\,.\label{eq:boundsmn}
\end{equation}
Remarkably, the bounds are only fulfilled in the case of $j^{2}=1/4$,
so that four of them saturate, corresponding to $p=\pm1/2$, and $p=\pm3/2$.
This is the case of Minkowski spacetime $\left(\mathcal{P}=-k/4\pi\right)$,
with four independent Killing vector-spinors that fulfill antiperiodic boundary conditions.
Hence, in spite of being maximally hypersymmetric, smooth solutions
whose energy is lower than the one of Minkowski spacetime are excluded
by the hypersymmetry bounds. 

It is worth highlighting that one arrives to similar conclusions in
the case of asymptotically flat spacetimes in supergravity \cite{BDMT}.
In fact, despite the analysis is fairly different, the supersymmetry
bounds precisely select the same spectrum, including the corresponding
ground states who saturate the bounds for spinors that fulfill different
periodicity conditions.

\section{Hypergravity reloaded \label{HypergravityReloaded}}

Let us look for a different theory of three-dimensional (hyper)gravity
that is still compatible with the asymptotically flat boundary conditions
described above, but now allowing the presence of spacetime torsion
even in vacuum. For simplicity, we consider modifications such that
the field equations are still of first order for the dreibein and
the spin connection. One interesting possibility is to include additional
terms, so that the action is given by 
\begin{equation}
I_{\gamma}=\frac{k}{4\pi}\int2R^{a}e_{a}+\gamma^{2}\epsilon_{abc}e^{a}e^{b}e^{c}+2\gamma T^{a}e_{a}\,.\label{eq:Igamma}
\end{equation}
Remarkably, despite the second (volume) term in the action looks like
a cosmological constant $\Lambda=-3\gamma^{2}$, the field equations
actually imply that the Riemann curvature vanishes. Indeed, the presence
of the last (parity-odd) term in the action has the effect of making
the volume term to act as the source for a fully antisymmetric torsion
in vacuum, being proportional to the volume element, given by 
\begin{equation}
T^{a}=-\gamma\varepsilon^{abc}e_{b}e_{c}\,,\label{eq:Torsion1}
\end{equation}
so that the remaining field equations fix the curvature two-form according
to
\begin{align}
R^{a} & =\frac{1}{2}\gamma^{2}\varepsilon^{abc}e_{b}e_{c}\,.\label{eq:Curvature1}
\end{align}
Therefore, eq. \eqref{eq:Torsion1} implies that the spin connection
splits as $\omega^{a}=\bar{\omega}^{a}+\kappa^{a}$, where $\bar{\omega}^{a}$
is the (torsionless) Levi-Civita connection, and the contorsion reads
$\kappa^{a}=-\gamma e^{a}$. The curvature two-form is then given by
$R^{a}=\bar{R}^{a}+\frac{1}{2}\gamma^{2}\varepsilon^{abc}e_{b}e_{c}$,
and hence, equation \eqref{eq:Curvature1} implies the vanishing of
the Riemann tensor, i. e., $\bar{R}_{a}=\frac{1}{4}\varepsilon_{a\rho\tau}R_{\quad\mu\nu}^{\rho\tau}dx^{\mu}dx^{\nu}=0$.

The most general theory that possesses the features described above
is obtained by considering the addition of the Lorentz-Chern-Simons
form, $L(\omega)=\omega^{a}d\omega_{a}+\frac{1}{3}\varepsilon_{abc}\omega^{a}\omega^{b}\omega^{c}$,
with an independent coupling $\mu$, provided the remaining couplings
in \eqref{eq:Igamma} are suitable shifted. The action is then given
by
\begin{equation}
I_{\mu,\gamma}=\frac{k}{4\pi}\int2\left(1+\mu\gamma\right)R^{a}e_{a}+\gamma^{2}\left(1+\mu\frac{\gamma}{3}\right)\epsilon_{abc}e^{a}e^{b}e^{c}+\mu L(\omega)+\gamma\left(2+\mu\gamma\right)T^{a}e_{a}\ .\label{eq:Imugamma}
\end{equation}
Noteworthy, despite the fact that the Lorentz-Chern-Simons form is
not a boundary term, the shifts in the other couplings are such that
the field equations in vacuum just become reshuffled, coinciding with
the previous ones for $\mu=0$, given by \eqref{eq:Torsion1} and
\eqref{eq:Curvature1}. Actually, one should highlight that both actions,
\eqref{eq:Igamma} and \eqref{eq:Imugamma}, differ off-shell, which
reflects through the fact that the canonical generators do not have
the same form. Consequently, as in the case of supergravity \cite{BDMT},
the asymptotic symmetry algebra of the latter acquires an additional
central extension with respect to the former (see below).

The locally hypersymmetric extension of the theory described by \eqref{eq:Imugamma}
is given by the following action 
\begin{equation}
I_{\mu,\gamma,\psi}=I_{\mu,\gamma}+\frac{k}{4\pi}\int i\bar{\psi}_{a}\left(D+\frac{3}{2}\gamma e^{b}\Gamma_{b}\right)\psi^{a}\,,\label{eq:IHReloaded}
\end{equation}
which is invariant under the following local hypersymmetry transformations:
\begin{equation}
\delta e^{a}=\frac{3}{2}i\bar{\epsilon}_{b}\Gamma^{a}\psi^{b}\quad,\quad\delta\omega^{a}=-\frac{3}{2}i\gamma\bar{\epsilon}_{b}\Gamma^{a}\psi^{b}\quad,\quad\delta\psi^{a}=D\epsilon^{a}+\frac{3}{2} \gamma e^{b}\Gamma_{b}\epsilon^{a}-\gamma  e_{b} \Gamma^{a}\epsilon^{b}\,.\label{eq:TransfReloaded}
\end{equation}
The field equations now read
\begin{equation*}
R^{a}=\frac{1}{2}\gamma^{2}\varepsilon^{abc}e_{b}e_{c}-\frac{3}{4}i\gamma\bar{\psi}_{b}\Gamma^{a}\psi^{b}\quad,\quad T^{a}=-\gamma\varepsilon^{abc}e_{b}e_{c}+\frac{3}{4}i\bar{\psi}_{b}\Gamma^{a}\psi^{b},
\end{equation*}
\begin{equation}
 D\psi^{a}=-\frac{3}{2}\gamma e^{b}\Gamma_{b}\psi^{a}+\gamma e_{b} \Gamma^{a}\psi^{b}\,.\label{eq:FeqsReloaded}
\end{equation}
Note that in the case of $\mu=\gamma=0$, the action \eqref{eq:IHReloaded},
the transformations rules \eqref{eq:TransfReloaded}, and the field
equations \eqref{eq:FeqsReloaded}, reduce to the ones of the locally
hypersymmetric extension of General Relativity, given by eqs. \eqref{eq:Ihyper},
\eqref{eq:TL}, and \eqref{eq:FE}, respectively.

As outlined in \cite{FMT-3D-HYGRA}, in analogy with the case of supergravity
\cite{GTW-Reloaded}, \cite{BDMT}, the action \eqref{eq:Imugamma}
can be formulated as a Chern-Simons one for the hyper-Poincaré group
\eqref{eq:hyperPoincare} by virtue of a simple modification of the
invariant bilinear form, and a suitable shift of the spin connection.
Indeed, the invariant bilinear form in \eqref{eq:IBF} can be consistently
modified to admit an additional nonvanishing component given by
\begin{equation}
\langle J_{a},J_{b}\rangle=\mu\eta_{ab}\,,\label{eq:JJmu}
\end{equation}
so that the Chern-Simons action \eqref{eq:Chern-Simons} now depends
on a different hyper-Poincaré-algebra-valued gauge field, defined
as
\begin{equation}
A=e^{a}P_{a}+\hat{\omega}^{a}J_{a}+\psi_{a}^{\alpha}Q_{\alpha}^{a}\ ,\label{A reloaded}
\end{equation}
with $\hat{\omega}^{a}:=\omega^{a}+\gamma e^{a}$. Therefore, in terms
of the covariant derivative with respect to $\hat{\omega}^{a}$ and
its corresponding curvature, given by $\hat{D}$ and $\hat{R}^{a}$,
respectively, up to a surface term, the Chern-Simons action reduces
to
\begin{equation}
I_{\mu,\gamma,\psi}=\frac{k}{4\pi}\int2\hat{R}^{a}e_{a}+\mu L(\hat{\omega})+i\bar{\psi}_{a}\hat{D}\psi^{a}\ ,
\end{equation}
which precisely agrees with \eqref{eq:IHReloaded}. Note that the
field equations \eqref{eq:FeqsReloaded} correspond to the vanishing
of the components of the curvature associated to \eqref{A reloaded},
so that they can be compactly written as $F=dA+A^{2}=0$, being manifestly
covariant under the full hyper-Poincaré group. 

One of the advantages of having formulated the extension of hypergravity
with parity-odd terms as a Chern-Simons theory, is that its asymptotically
flat structure can be directly obtained along the lines of the results
in section \ref{Asymptotics}. 

The asymptotically flat boundary conditions for the connection \eqref{A reloaded}
are then proposed to be precisely as in eqs. \eqref{eq:Aconh}, \eqref{eq:aphi},
and \eqref{eq:au}, so that the asymptotic fall-off of the spin connection
$\omega^{a}$ becomes modified. Therefore, the asymptotic symmetries
remain the same as in section \ref{Asymptotics}, being spanned by
the hyper-Poincaré algebra valued parameter $\lambda=\lambda\left[T,Y,\mathcal{E}\right]$
given by \eqref{eq:lambda}. The global charges are then found to
acquire a correction due to the additional component of the invariant
bilinear form in \eqref{eq:JJmu}, so that they now read 
\begin{equation}
Q\left[T,Y,\mathcal{E}\right]=-\int\left(T\mathcal{P}+Y\tilde{\mathcal{J}}-i\mathcal{E}\psi\right)d\phi\,,\label{eq:Qtilde-reloaded}
\end{equation}
with $\mbox{\ensuremath{\tilde{\mathcal{J}}}=\ensuremath{\mathcal{J}}+\ensuremath{\mu\mathcal{P}}}$,
and do not depend on the parameter $\gamma$. Note that the shift
in the canonical generator associated to $Y$ implies that in the
extended theory, even static configurations, as it is the case of
Minkowski spacetime, may carry angular momentum.

It is then simple to verify that, once the canonical generators are
expanded in modes, their nonvanishing Poisson brackets are given by
\begin{eqnarray}
i\left\{ \tilde{\mathcal{J}}_{m},\mathcal{\tilde{\mathcal{J}}}_{n}\right\}  & = & \left(m-n\right)\mathcal{\tilde{\mathcal{J}}}_{m+n}+\mu km^{3}\delta_{m+n,0}\,,\nonumber \\
i\left\{ \tilde{\mathcal{J}}_{m},\mathcal{P}_{n}\right\}  & = & \left(m-n\right)\mathcal{P}_{m+n}+km^{3}\delta_{m+n,0}\,,\nonumber \\
i\left\{ \mathcal{\tilde{\mathcal{J}}}_{m},\mathcal{\psi}_{n}\right\}  & = & \left(\frac{3m}{2}-n\right)\psi_{m+n}\,,\label{eq:hyperBMS-2-1}\\
i\left\{ \psi_{m},\psi_{n}\right\}  & = & \frac{1}{2}\left(3m^{2}-4mn+3n^{2}\right)\mathcal{P}_{m+n}+\frac{9}{4k}\sum_{q}\mathcal{P}_{m+n-q}\mathcal{P}_{q}+km^{4}\delta_{m+n,0}\,,\nonumber 
\end{eqnarray}
which corresponds to a hypersymmetric extension of the BMS$_{3}$
algebra, with an additional central extension along its Virasoro subalgebra.

\section{General Relativity minimally coupled to half-integer spin fields
\label{HypergravityS}}

In the generic case of fermionic fields of spin $n+\frac{3}{2}$,
and in the absence of cosmological constant, the hypergravity action
reads \cite{AD-3D-HYGRA}, \cite{FMT-3D-HYGRA} 
\begin{equation}
I=\frac{k}{4\pi}\int2R^{a}e_{a}+i\bar{\psi}_{a_{1}\dots a_{n}}D\psi^{a_{1}\dots a_{n}}\,,\label{eq:ActionS-2}
\end{equation}
where $\psi_{a_{1}\dots a_{n}}$ describes a Grassmann-valued 1-form
that is $\Gamma-$traceless, i. e., $\Gamma^{a_{1}}\psi_{a_{1}\dots a_{n}}=0$,
and completely symmetric in its vector indices. Its covariant derivative
can be conveniently written as
\begin{equation}
D\psi^{a_{1}\dots a_{n}}=d\psi^{a_{1}\dots a_{n}}+\left(n+\frac{1}{2}\right)\omega^{b}\Gamma_{b}\psi^{a_{1}\dots a_{n}}-\omega_{b}\Gamma^{(a_{1}}\psi^{a_{2}\dots a_{n})b}\,.
\end{equation}
The standard supergravity action in \cite{Deser-Kay}, \cite{Deser-CTS},
\cite{Marcus-Schwarz} is then recovered for $n=0$, while the theory
discussed in section \ref{hypergravity5/2} corresponds to $n=1$.

The generic theory can also be formulated in terms of a Chern-Simons
action for a gauge field that takes values in the hyper-Poincaré algebra,
given by
\begin{equation}
A=e^{a}P_{a}+\omega^{a}J_{a}+\psi_{a_{1}\dots a_{n}}^{\alpha}Q_{\alpha}^{a_{1}\dots a_{n}}\,.\label{eq:Connection-n-1}
\end{equation}
Here $Q_{\alpha}^{a_{1}\dots a_{n}}$ correspond to $\Gamma-$traceless
fermionic generators of spin $n+\frac{1}{2}$. The explicit expression
of the generic hyper-Poincaré algebra can be compactly written in
terms of its Maurer-Cartan form (see appendix \ref{hyperPoincareSpinS}).
The field equations then read $F=dA+A^{2}=0$, with
\begin{equation}
F=R^{a}J_{a}+\tilde{T}^{a}P_{a}+D\psi_{a_{1}\dots a_{n}}^{\alpha}Q_{\alpha}^{a_{1}\dots a_{n}}\,,\label{eq:Curvature-n-1}
\end{equation}
where the hypercovariant torsion is now given by
\begin{equation}
\tilde{T}^{a}=T^{a}-\frac{1}{2}\left(n+\frac{1}{2}\right)i\bar{\psi}_{a_{1}\dots a_{n}}\Gamma^{a}\psi^{a_{1}\dots a_{n}}\,.
\end{equation}
Thus, by construction, the action is invariant under gauge transformations
generated by $\lambda=\epsilon_{a_{1}\dots a_{n}}^{\alpha}Q_{\alpha}^{a_{1}\dots a_{n}}$,
so that
\begin{eqnarray}
\delta e^{a} & = & \left(n+\frac{1}{2}\right)i\bar{\epsilon}_{a_{1}\dots a_{n}}\Gamma^{a}\psi^{a_{1}\dots a_{n}}\,,\nonumber \\
\delta\omega^{a} & = & 0\,,\label{eq:transN}\\
\delta\psi^{a_{1}\dots a_{n}} & = & D\epsilon^{a_{1}\dots a_{n}}\,.\nonumber 
\end{eqnarray}

\subsection{Killing tensor-spinors} \label{Killing tensor-spinors}

According to \eqref{eq:transN}, a purely bosonic configuration is
invariant under local hypersymmetry transformations provided the following
``Killing tensor-spinor equation'' is fulfilled:
\begin{equation}
d\epsilon_{a_{1}\dots a_{n}}+\left(n+\frac{1}{2}\right)\omega^{b}\Gamma_{b}\epsilon_{a_{1}\dots a_{n}}-\omega^{b}\Gamma_{(a_{1}}\epsilon_{a_{2}\dots a_{n})b}=0\,.\label{eq:KillingT-Seq}
\end{equation}
Since the field equations imply the vanishing of the curvature two-form
$R^{a}$, the general solution of \eqref{eq:KillingT-Seq} is now
given by
\begin{equation}
\epsilon_{a_{1}\cdots a_{n}}^{\alpha}=\left(g_{S}^{-1}\right)_{\beta}^{\alpha}\left(g_{V}\right)_{a_{1}}^{b_{1}}\cdots\left(g_{V}\right)_{a_{n}}^{b_{n}}\eta_{b_{1}\cdots b_{n}}^{\beta}\,,\label{eq:Killing T-S}
\end{equation}
where $g_{S}$ and $g_{V}$ are defined in eq. \eqref{eq:GroupElements}.
As explained in section \ref{Section3}, both stand for the same group
element $g$ that determines the spin connection, $\omega=g^{-1}dg$,
but expressed in the spinor and the vector representations, respectively.
In the generic case, $\eta_{b_{1}\cdots b_{n}}^{\beta}$ is a constant
$\Gamma-$traceless tensor-spinor. Unbroken hypersymmetries then correspond
to Killing tensor-spinors of the form \eqref{eq:Killing T-S}, that
are globally well-defined.

The hypersymmetry properties of the class of solutions discussed in
section \ref{Cosmology}, describing cosmological spacetimes and configurations
with conical defects, then go as follows. For any configuration with
$\mathcal{P}\neq0$, $g_{S}$ and $g_{V}$ are given by \eqref{eq:gsPneq0}
and \eqref{eq:gVPneq0}, respectively; while in the case of $\mathcal{P}=0$,
they read as in eq. \eqref{eq:gP0}. Therefore, in the case of $\mathcal{P}>0$
the solutions cannot possess globally-defined Killing tensor-spinors,
because $\epsilon_{a_{1}\cdots a_{n}}^{\alpha}$ in \eqref{eq:Killing T-S}
do not fulfill neither periodic nor antiperiodic boundary conditions.
This means that hypersymmetries are necessarily broken for cosmological
spacetimes. 

By virtue of \eqref{eq:Killing T-S} and \eqref{eq:gP0}, configurations
with $\mathcal{P}=0$ only admit constant Killing tensor-spinors that
fulfill the following condition:
\begin{equation}
\left(n+\frac{1}{2}\right)\Gamma_{1}\epsilon_{a_{1}\dots a_{n}}-\Gamma_{(a_{1}}\epsilon_{a_{2}\dots a_{n})1}=0\,,
\end{equation}
which implies that they have a single nonvanishing component given
by $\epsilon_{00\cdots0}^{-}=\eta_{00\cdots0}^{-}$. Therefore, this
class of spacetimes possesses just one unbroken hypersymmetry, which
relates to the fact that there is only one hypersymmetry bound that
saturates for fermions with periodic boundary conditions (see below).

As explained in section \ref{Cosmology}, smooth solutions with conical
defects are maximally hypersymmetric and their energy is determined
by $\mathcal{P}=-kj^{2}/\pi<0$, where $j$ is a (half-)integer. For
this class of configurations, the explicit form of the Killing tensor-spinors
is then given by \eqref{eq:Killing T-S}, with $g_{S}$ and $g_{V}$
being described by eqs. \eqref{eq:gsP<0} and \eqref{eq:gVP<0}, respectively.
It will also be shown below that conical surpluses are excluded by
the hypersymmetry bounds, which are fulfilled only for $j^{2}=1/4$,
which corresponds to the case of Minkowski spacetime.

\subsection{Asymptotically flat structure and hypersymmetry bounds}

In order to describe the asymptotically flat behaviour of hypergravity
in the generic case, it is convenient to make use of the $\Gamma-$traceless
condition of the fields and the generators, which amounts to reduce
the number of independent components. The hyper-Poincaré algebra in
\eqref{eq:MC-algebra} can then be alternatively written as
\begin{eqnarray}
\left[\hat{\mathcal{J}}_{m},\hat{\mathcal{J}}_{n}\right] & = & \left(m-n\right)\hat{\mathcal{J}}_{m+n}\nonumber \\
\left[\hat{\mathcal{J}}_{m},\hat{\mathcal{P}}_{n}\right] & = & \left(m-n\right)\hat{\mathcal{P}}_{m+n}\,,\nonumber \\
\left[\hat{\mathcal{J}}_{m},\hat{\mathcal{Q}}_{p}\right] & = & \left(sm-p\right)\hat{\mathcal{Q}}_{m+p}\,,\\
\left\{ \hat{\mathcal{Q}}_{p},\hat{\mathcal{Q}}_{q}\right\}  & = & f_{p,q}^{\left(s\right)}\hat{\mathcal{P}}_{p+q}\,,\nonumber 
\end{eqnarray}
with $m,n=0,\pm1$, and $p,q=\pm\frac{1}{2},\pm\frac{3}{2}\dots,\pm s$,
where $s$ stands for the spin of the fermionic generators $\hat{\mathcal{Q}}_{p}$.
The structure constants fulfill $f_{p,q}^{\left(s\right)}=f_{q,p}^{\left(s\right)}=f_{-p,-q}^{\left(s\right)}$,
and the nonvanishing ones are given by
\begin{equation}
f_{p,-p}^{\left(s\right)}=-\frac{2p}{s+p+1}f_{p,-p-1}^{\left(s\right)}=\left(-1\right)^{p+\frac{1}{2}}2p\prod_{k=\frac{1}{2}}^{|p|}\frac{\left(2s+2k\right)}{\left(2s-2\left(k-1\right)\right)}\text{\,,}
\end{equation}
provided $|p+q|\leq1$. Here the fermionic generators have been normalized
according to $f_{\frac{1}{2},-\frac{1}{2}}^{\left(s\right)}=-1$.

It is amusing to verify that the Jacobi identity now translates into
the fact that the structure constants $f_{p,q}^{\left(s\right)}$
solve the following recursion relation:
\begin{equation}
\left(m-\left(p+q\right)\right)f_{q,p}^{\left(s\right)}-\left(sm-p\right)f_{q,m+p}^{\left(s\right)}-\left(sm-q\right)f_{p,m+q}^{\left(s\right)}=0\,.
\end{equation}
For later purposes it is useful to note that
\begin{equation}
f_{s,-s}^{\left(s\right)}=\left(-1\right)^{s+\frac{1}{2}}\frac{2s}{2s+1}\frac{\left(4s\right)!!}{\left(2s-1\right)!!^{2}}\,.\label{eq:fs-s}
\end{equation}
The structure constants can also be conveniently written as 
\begin{equation}
f_{m,n}^{\left(s\right)}=\sum_{l=0}^{s-\frac{1}{2}}h_{m,n}^{(l)}\,,\label{eq:fmn-hmn}
\end{equation}
where $h_{m,n}^{(l)}$ stand for homogeneous polynomials of degree
$2l$ in $m$, $n$, i. e., $h_{\lambda m,\lambda n}^{(l)}=\lambda^{2l}h_{m,n}^{(l)}$.
Indeed, as it is shown below, the asymptotic symmetry algebra can
be naturally expressed in terms of $h_{m,n}^{(l)}$, where $m$, $n$
are extended to be arbitrary (half-)integers. Note that in the case
of supergravity $f_{m,n}^{\left(1/2\right)}=-1$, while for fermionic
generators of spin $s=3/2$, the form of $f_{m,n}^{\left(3/2\right)}$
can be read from eq. \eqref{eq:hyperPoincareModos}. In the case of
fermionic generators with $s=5/2$, $7/2$ the explicit form of $f_{m,n}^{\left(5/2\right)}$
and $f_{m,n}^{\left(7/2\right)}$ is given in appendices \ref{Spin5/2}
and \ref{Spin7/2}, respectively.

Following the lines of section \ref{Asymptotics}, the asymptotic
form of the gauge field can be written as in eq. \eqref{eq:Aconh},
so that at a fixed time slice, the dynamical field is proposed to
be given by
\begin{equation}
a_{\phi}=\hat{\mathcal{J}}_{1}-\frac{\pi}{k}\left(\mathcal{J}\hat{\mathcal{P}}_{-1}+\mathcal{P}\hat{\mathcal{J}}_{-1}+\alpha_{s}\psi\hat{\mathcal{Q}}_{-s}\right)\,,\label{eq:aphi-S}
\end{equation}
with
\begin{equation}
\alpha_{s}=\left(f_{-s,s+1}^{\left(s\right)}\right)^{-1}=-2s\left(f_{s,-s}^{\left(s\right)}\right)^{-1}\,,
\end{equation}
and $f_{s,-s}^{\left(s\right)}$ can be read from eq. \eqref{eq:fs-s}.

The asymptotic symmetries are then generically spanned by a hyper-Poincaré-valued
parameter of the form
\begin{equation}
\lambda=T\hat{\mathcal{P}}_{1}+Y\hat{\mathcal{J}}_{1}+\mathcal{E}\hat{\mathcal{Q}}_{s}+\eta_{\left(s\right)}\left[T,Y,\mathcal{E}\right]\,,\label{eq:lambdaS}
\end{equation}
where $\eta_{\left(s\right)}\left[T,Y,\mathcal{E}\right]$ goes along
all but the lowest weight generators, provided the fields $\mathcal{J}$,
$\mathcal{P}$, $\psi$ transform in a suitable way. 

The asymptotic form of the Lagrange multiplier can then be written
in terms of the chemical potentials according to 
\begin{equation}
a_{u}=\lambda\left[\mu_{\mathcal{P}},\mu_{\mathcal{J}},\mu_{\mathcal{\psi}}\right]\,.\label{eq:auS}
\end{equation}
Its form is preserved under evolution in time as long as the field
equations are fulfilled in the asymptotic region, and the parameters
are subject to appropriate conditions, being described by first order
equations in time. 

In order to integrate the variation of the canonical generators in
\eqref{eq:CS-Charge}, one needs the relevant fermionic component
of the invariant bilinear form, which is given by 
\begin{equation}
\left\langle \hat{\mathcal{Q}}_{s},\hat{\mathcal{Q}}_{-s}\right\rangle =2\alpha_{s}^{-1}\,,
\end{equation}
so that the global charges in the generic case acquire the same form
as in eq. \eqref{eq:Charge5/2}, i. e.,
\begin{equation}
Q\left[T,Y,\mathcal{E}\right]=-\int\left(T\mathcal{P}+Y\mathcal{J}-i\mathcal{E}\psi\right)d\phi\,.
\end{equation}
Once expanded in Fourier modes, the nonvanishing Poisson brackets
of the canonical generators are given by
\begin{eqnarray}
i\left\{ \mathcal{J}_{m},\mathcal{J}_{n}\right\}  & = & \left(m-n\right)\mathcal{J}_{m+n}\,,\nonumber \\
i\left\{ \mathcal{J}_{m},\mathcal{P}_{n}\right\}  & = & \left(m-n\right)\mathcal{P}_{m+n}+km^{3}\delta_{m+n,0}\,,\nonumber \\
i\left\{ \mathcal{J}_{m},\psi_{n}\right\}  & = & \left(sm-n\right)\psi_{m+n}\,,\label{eq:ASA-S}\\
i\left\{ \psi_{m},\psi_{n}\right\}  & = & \sum_{q=0}^{s-1/2}\frac{\left(-1\right)^{2s-q}}{s-q+\frac{1}{2}}\left(\frac{2}{k}\right)^{s-q-\frac{1}{2}}h_{m,n}^{(q)}\mathcal{P}_{m+n}^{s-q-\frac{1}{2}}+\left(-1\right)^{s-\frac{1}{2}}\frac{2km^{2s+1}}{\alpha_{s}\left(2s\right)!}\delta_{m+n,0}+\Xi_{m+n}^{\left(s\right)}\,.\nonumber 
\end{eqnarray}
The conformal weight of the fermionic generators $\psi_{n}$ with
respect to $\mathcal{J}_{m}$ is given by $\Delta=s+1$. Here $h_{m,n}^{(q)}$
stand for the homogeneous polynomials defined through eq. \eqref{eq:fmn-hmn},
extended to the case of (half-)integers, and 
\begin{equation}
\mathcal{P}_{m+n}^{r}:=\sum_{i_{1},\cdots i_{r}}\mathcal{P}_{m+n-i_{1}\dots-i_{r}}\mathcal{P}_{i_{1}}\cdots\mathcal{P}_{i_{r}}\,.
\end{equation}
Here $\Xi_{m+n}^{\left(s\right)}$ stands for the mode expansion of
nonlinear terms that contains derivatives of $\mathcal{P}$, and becomes
nontrivial provided $s>3/2$. Indeed, according to eqs. \eqref{eq:superBMS}
and \eqref{eq:hyperBMS}, in the case of supergravity $\left(s=1/2\right)$,
and for $s=3/2$, one finds that $\Xi_{m+n}^{\left(1/2\right)}=\Xi_{m+n}^{\left(3/2\right)}=0$;
while for $s=5/2$ it is proportional to the mode expansion of $\left(\mathcal{P}\mbox{\ensuremath{'}}\right)^{2}$
(see eq. \eqref{eq:Xi52}). The explicit form of $\Xi^{\left(7/2\right)}$
is given in eq. \eqref{eq:Xi7/2}.

As in section \ref{Bounds}, the asymptotic symmetry algebra \eqref{eq:ASA-S}
also implies the existence of nonlinear bounds for the energy. Indeed,
making the same assumptions, for $m=-n=p$, the (fermionic) anticommutator
is manifestly positive-definite. Furthermore, since in the ``rest
frame'' the bosonic global charges just correspond to $\mathcal{P}_{0}$,
the nonlinear terms described by $\Xi_{m+n}^{\left(s\right)}$ in
the fermionic anticommutator do not contribute. Therefore, in the
generic case the bounds are given by
\begin{equation}
\prod_{i=0}^{n}\left(p^{2}+\left(2i+1\right)^{2}\frac{\pi\mathcal{P}}{k}\right)\geq0\,,\label{eq:Bounds-S}
\end{equation}
where $p$ is a (half-)integer for the case of fermionic fields of
spin $s=n+3/2$ that fulfill (anti)periodic boundary conditions.

It is then clear that the bounds are fulfilled for configurations
with $\mathcal{P}>0$, as it is the case of cosmological spacetimes.
The fact that they never saturate agrees with the nonexistence of
globally-defined Killing tensor-spinors. Note that in the case of
$\mathcal{P}=0$ the bounds are also satisfied, while the one with
$p=0$ saturates, which corresponds to the fact that configurations
of this sort admit a single unbroken hypersymmetry, being generated
by a constant Killing tensor-spinor.

For the class of maximally hypersymmetric smooth solutions with negative
energy $\left(\mathcal{P}=-kj^{2}/\pi\right)$ described in section
\ref{Cosmology}, the bounds \eqref{eq:Bounds-S} read
\begin{equation} \label{eq:boundspin-s}
\prod_{i=0}^{n}\left(p^{2}-\left(2i+1\right)^{2}j^{2}\right)\geq0\,,
\end{equation}
which implies that the only case that fulfills all of them, also saturate
the ones for $p=\pm(2i+1)/2$, with $i=0,1,\dots,n$, and corresponds
to $j^{2}=1/4$. Thus, Minkowski spacetime becomes naturally selected
at the ground state in the case of fermions that satisfy antiperiodic
boundary conditions, possessing the maximum number of Killing tensor-spinors
described by \eqref{eq:Killing T-S}, with \eqref{eq:gsP<0} and \eqref{eq:gVP<0}.

In sum, in the case of fermions that fulfill periodic boundary conditions
the energy spectrum is nonnegative ($\mathcal{P}\geq0$), so that
the allowed class of solutions is generically characterized by the
cosmological spacetimes described in section \ref{Cosmology}. The
ground state is then given by a configuration of vanishing energy
that saturates only one of the bounds ($p=0$). This corresponds to
the null orbifold which, as shown in section \ref{Killing tensor-spinors},
possesses a single Killing tensor-spinor. If the fermions satisfy
antiperiodic boundary conditions, the spectrum becomes enlarged since
the bounds now imply that $\mathcal{P}\geq-k/4\pi$. Nonetheless,
since conical defects and surpluses generically do not fulfill the
field equation in vacuum, they are discarded unless they are smooth.
According to \eqref{eq:boundspin-s}, in this case the ground state
saturates as many bounds as the maximum number of Killing tensor-spinors,
and it can be identified with Minkowski spacetime, so that the spectrum
acquires a gap.

\section{Final remarks \label{FinalRemarks}}

In the case of fermionic fields of spin $s=n+3/2$, the locally hypersymmetric
extension of the action $I_{\mu,\gamma}$ in \eqref{eq:Imugamma},
that includes parity-odd terms, can also be formulated as a Chern-Simons
theory for the hyper-Poincaré group in \eqref{eq:MC-algebra}. In
order to carry out this task, the invariant bilinear form has to be
suitably modified, so that it acquires additional components being
determined by eq. \eqref{eq:JJmu}. The gauge field reads
\begin{equation}
A=e^{a}P_{a}+\hat{\omega}^{a}J_{a}+\psi_{a_{1}\dots a_{n}}^{\alpha}Q_{\alpha}^{a_{1}\dots a_{n}}\,,\label{eq:Connection-n-reloaded}
\end{equation}
where, as in section \ref{HypergravityReloaded}, $\hat{\omega}^{a}=\omega^{a}+\gamma e^{a}$.
Therefore, up to a boundary term, the action of the extended hypergravity
theory reduces to 
\begin{equation}
I_{\mu,\gamma,\psi_{n}}=I_{\mu,\gamma}+\frac{k}{4\pi}\int i\bar{\psi}_{a_{1}\dots a_{n}}\left[D+\left(n+\frac{1}{2}\right)\gamma e^{b}\Gamma_{b}\right]\psi^{a_{1}\dots a_{n}}\,,\label{eq:IHReloaded-1-1}
\end{equation}
being by construction locally invariant under
\begin{eqnarray}
\delta e^{a} & = & \left(n+\frac{1}{2}\right)i\bar{\epsilon}_{a_{1}\dots a_{n}}\Gamma^{a}\psi^{a_{1}\dots a_{n}}\,,\nonumber \\
\delta\omega^{a} & = & -\left(n+\frac{1}{2}\right)i\gamma\bar{\epsilon}_{a_{1}\dots a_{n}}\Gamma^{a}\psi^{a_{1}\dots a_{n}}\,,\label{eq:transN-reloaded-1}\\
\delta\psi^{a_{1}\dots a_{n}} & = & \left[D+\left(n+\frac{1}{2}\right)\gamma e^{b}\Gamma_{b}\right]\epsilon^{a_{1}\dots a_{n}}-\gamma e_{b}\Gamma^{(a_{1}}\epsilon^{a_{2}\dots a_{n})b}\,.\nonumber 
\end{eqnarray}
Note that the extended hypergravity action \eqref{eq:IHReloaded-1-1},
and its corresponding local hypersymmetry transformations \eqref{eq:transN-reloaded-1},
agree with the corresponding ones for the locally hypersymmetric extension
of General Relativity, given by \eqref{eq:ActionS-2} and \eqref{eq:transN},
respectively, in the case of $\mu=\gamma=0$. Consequently, a suitable
set of asymptotically flat boundary conditions for the extended theory
is also proposed to be described by gauge fields of the form \eqref{eq:Aconh},
\eqref{eq:aphi-S}, and \eqref{eq:auS}. The canonical generators
of the asymptotic symmetries then reduce to the ones in eq. \eqref{eq:Qtilde-reloaded},
with $\mbox{\ensuremath{\tilde{\mathcal{J}}}=\ensuremath{\mathcal{J}}+\ensuremath{\mu\mathcal{P}}}$,
so that their algebra is readily found to be described by \eqref{eq:ASA-S},
but with an additional central extension along the Virasoro subalgebra,
precisely as in eq. \eqref{eq:hyperBMS-2-1}. 

It is worth pointing out that prescribing the asymptotic behaviour
of gauge fields to be described by deviations with respect to a reference
background that go along the highest weight generators of the algebra,
turns out to be a very successful strategy. Indeed, this is not only
the case of General Relativity in three spacetime dimensions \cite{CH-vanDriel},
but it is also so for its locally supersymmetric extension with or
without cosmological constant \cite{Henneaux-Maoz-Schwimmer}, \cite{BDMT},
or even when the theory is nonminimally coupled to higher spin fields
\cite{Henneaux-Rey}, \cite{CFPT-higher spin}, \cite{HLPR-super higher spin},
\cite{ABFGR-HSF3D}, \cite{GMPT-HSF3D}, \cite{Henneaux-CP3D}, \cite{Gutperle-Hijano-Samani},
\cite{BHPTT-GBH3D}, \cite{Gary-HSChP}, \cite{Matulich-HSChP}, \cite{Gutperle-Li}.

One of the interesting features of dealing with hypersymmetry, is
that nonlinear bounds for the energy have been shown to naturally
emerge from the anticommutator of fermionic generators. In the case
of vanishing cosmological constant, the hypersymmetry bounds for the
theory in vacuum turn out to exclude solutions that describe conical
defects and surpluses \cite{DJT-3D}, \cite{Deser-Jackiw}, despite
the latter are maximally (hyper)symmetric. In presence of higher spin
fields, the analogue of this class of configurations has been discussed
in \cite{CGGR-conical defects}, \cite{Datta-David}, \cite{CLW},
\cite{CPR-conical solutions}, \cite{CF-conical defects}, \cite{LLW-modular},
\cite{Raeymaekers}, \cite{Hyper-AdS}. It is then worth highlighting
that, according to the results that have been recently obtained for
hypergravity with negative cosmological constant \cite{Hyper-AdS},
one is naturally led to expect that only a suitable subset of asymptotically
flat solitonic-like solutions might fulfill the hypersymmetry bounds,
for which the higher spin charges become tuned in terms of the mass.
Indeed, it is amusing to verify that the gauge group $Sp\left(4\right)\otimes OSp\left(1|4\right)$
admits an inequivalent Inönü-Wigner contraction as compared with the
one described in section \ref{FlatLimit}, so that the electric-like
spin-4 charges cannot be consistently decoupled in this alternative
flat limit of the $\left(0,1\right)$ theory. This contraction is
defined through a different change of basis:%
\footnote{Note that $OSp\left(1|4\right)$, corresponding to the super-AdS$_{4}$
group, as well as the superconformal group in three spacetime dimensions,
admits two interesting consistent ``flat limits'' $\left(\ell\rightarrow\infty\right)$,
which can be obtained rescaling the generators either as in eq. \eqref{eq:Contraction1},
or in \eqref{eq:Contraction2}, provided the left copy is switched
off. %
}

\[
\hat{\mathcal{P}}_{i}=\frac{1}{\ell}\left(\hat{\mathcal{L}}_{i}^{+}+\hat{\mathcal{L}}_{-i}^{-}\right)\quad,\quad\hat{\mathcal{J}}_{i}=\hat{\mathcal{L}}_{i}^{+}-\hat{\mathcal{L}}_{-i}^{-}\,,
\]
\begin{equation}
\mathcal{\hat{W}}_{n}=\frac{1}{\ell}\left(\hat{\mathcal{U}}_{n}^{+}+\hat{\mathcal{U}}_{-n}^{-}\right)\quad,\quad\hat{\mathcal{V}}_{n}=\hat{\mathcal{U}}_{n}^{+}-\hat{\mathcal{U}}_{-n}^{-}\quad,\quad\hat{\mathcal{Q}}_{p}=\sqrt{\frac{6}{\ell}}\hat{\mathcal{S}}_{p}^{+}\,,\label{eq:Contraction2}
\end{equation}
where $\hat{\mathcal{L}}_{i}^{-}$, $\mathcal{\hat{U}}_{n}^{-}$ stand
for the generators of the (left) $sp\left(4\right)$ algebra, and
$\hat{\mathcal{L}}_{i}^{+}$, $\hat{\mathcal{U}}_{n}^{+}$, $\hat{\mathcal{S}}_{p}^{+}$
span the (right) $osp\left(1|4\right)$ algebra. Thus, in the limit
of large AdS radius, $\ell\rightarrow\infty$, the nonvanishing components
of the (anti)commutators of the new algebra read 
\begin{eqnarray}
\left[\hat{\mathcal{J}}_{i},\hat{\mathcal{J}}_{j}\right] & = & \left(i-j\right)\hat{\mathcal{J}}_{i+j}\quad,\quad\left[\hat{\mathcal{J}}_{i},\hat{\mathcal{P}}_{j}\right]=\left(i-j\right)\hat{\mathcal{P}}_{i+j}\,,\nonumber \\
\left[\hat{\mathcal{J}}_{i},\hat{\mathcal{W}}_{n}\right] & = & \left(3i-n\right)\hat{\mathcal{W}}_{i+n}\quad,\quad\left[\hat{\mathcal{J}}_{i},\hat{\mathcal{V}}_{n}\right]=\left(3i-n\right)\hat{\mathcal{V}}_{i+n}\,,\nonumber \\
\left[\hat{\mathcal{P}}_{i},\hat{\mathcal{V}}_{n}\right] & = & \left(3i-n\right)\hat{\mathcal{W}}_{i+n}\quad,\quad\left[\hat{\mathcal{J}}_{i},\hat{\mathcal{Q}}_{p}\right]=\left(\frac{3i}{2}-p\right)\hat{\mathcal{Q}}_{i+p}\,,\nonumber\\
\left[\hat{\mathcal{V}}_{m},\hat{\mathcal{V}}_{n}\right] & = & \frac{1}{2^{2}3}\left(m-n\right)\left(\left(m^{2}+n^{2}-4\right)(m^{2}+n^{2}-\frac{2}{3}mn-9)-\frac{2}{3}\left(mn-6\right)mn\right)\hat{\mathcal{J}}_{m+n}\nonumber \\
 &  & +\frac{1}{6}\left(m-n\right)\left(m^{2}-mn+n^{2}-7\right)\hat{\mathcal{V}}_{m+n}\,,\label{eq:hyper-Poincare-4}\\
\left[\hat{\mathcal{V}}_{m},\hat{\mathcal{W}}_{n}\right] & = & \frac{1}{2^{2}3}\left(m-n\right)\left(\left(m^{2}+n^{2}-4\right)(m^{2}+n^{2}-\frac{2}{3}mn-9)-\frac{2}{3}\left(mn-6\right)mn\right)\hat{\mathcal{P}}_{m+n}\nonumber \\
 &  & +\frac{1}{6}\left(m-n\right)\left(m^{2}-mn+n^{2}-7\right)\hat{\mathcal{W}}_{m+n}\,,\nonumber\\
 \left[\hat{\mathcal{V}}_{m},\hat{\mathcal{Q}}_{p}\right] & = & \frac{1}{2^{3}3}\left(2m^{3}-8m^{2}p+20mp^{2}+82p-23m-40p^{3}\right)\hat{\mathcal{Q}}_{m+p}\,,\nonumber 
\end{eqnarray}

\begin{eqnarray}
\left\{ \hat{\mathcal{Q}}_{p},\hat{\mathcal{Q}}_{q}\right\}  & = & 3\hat{\mathcal{W}}_{p+q}+\frac{1}{2^{2}}\left(6p^{2}-8pq+6q^{2}-9\right)\hat{\mathcal{P}}_{p+q}\,,\nonumber 
\end{eqnarray}
which is to be regarded to span the gauge group of flat hypergravity
coupled to spin-4 fields. Here, $i,j=0,\pm1$, $m,n=0,\pm1,\pm2,\pm3$,
and $p,q=\pm\frac{1}{2},\pm\frac{3}{2}$. The mode expansion of the
asymptotically flat symmetry algebra of hypergravity with a fermionic
spin-$5/2$ field, being coupled to spin-4 fields, is then expected
to be such that the nonvanishing Poisson brackets are given by
\begin{eqnarray}
i\left\{ \mathcal{J}_{m},\mathcal{J}_{n}\right\}  & = & \left(m-n\right)\mathcal{J}_{m+n}\quad,\quad i\left\{ \mathcal{J}_{m},\mathcal{P}_{n}\right\} =\left(m-n\right)\mathcal{P}_{m+n}+km^{3}\delta_{m+n,0}\,,\nonumber \\
i\left\{ \mathcal{J}_{m},\mathcal{W}_{n}\right\}  & = & \left(3m-n\right)\mathcal{W}_{m+n}\quad,\quad i\left\{ \mathcal{J}_{m},\mathcal{V}_{n}\right\} =\left(3m-n\right)\mathcal{V}_{m+n}\,,\nonumber \\
i\left\{ \mathcal{P}_{m},\mathcal{V}_{n}\right\}  & = & \left(3m-n\right)\mathcal{W}_{m+n}\quad,\quad i\left\{ \mathcal{J}_{m},\psi_{n}\right\} =\left(\frac{3m}{2}-n\right)\psi_{m+n}\,,\nonumber \\
i\left\{ \mathcal{V}_{m},\mathcal{V}_{n}\right\}  & = & \frac{1}{2^{2}3^{2}}\left(m-n\right)\left(3m^{4}-2m^{3}n+4m^{2}n^{2}-2mn^{3}+3n^{4}\right)\mathcal{J}_{m+n}\nonumber \\
 &  & +\frac{1}{6}\left(m-n\right)\left(m^{2}-mn+n^{2}\right)\mathcal{V}_{m+n}-\frac{2^{3}3\pi}{k}\left(m-n\right)\Theta_{m+n}^{\left(6\right)}\nonumber \\
 &  & -\frac{7^{2}\pi}{3^{2}k}\left(m-n\right)\left(m^{2}+4mn+n^{2}\right)\Theta_{m+n}^{\left(4\right)}\nonumber \\
i\left\{ \mathcal{V}_{m},\mathcal{W}_{n}\right\}  & = & \frac{1}{2^{2}3^{2}}\left(m-n\right)\left(3m^{4}-2m^{3}n+4m^{2}n^{2}-2mn^{3}+3n^{4}\right)\mathcal{P}_{m+n}\label{eq:hyper-BMS-4}\\
 &  & +\frac{1}{6}\left(m-n\right)\left(m^{2}-mn+n^{2}\right)\mathcal{W}_{m+n}-\frac{2^{3}3\pi}{k}\left(m-n\right)\Omega_{m+n}^{\left(6\right)}\nonumber \\
 &  & -\frac{7^{2}\pi}{3^{2}k}\left(m-n\right)\left(m^{2}+4mn+n^{2}\right)\Omega_{m+n}^{\left(4\right)}+\frac{k}{2^{2}3^{2}}m^{7}\delta_{m+n,0}\,,\nonumber \\
i\left\{ \mathcal{V}_{m},\psi_{n}\right\}  & = & \frac{1}{2^{2}3}\left(m^{3}-4m^{2}n+10mn^{2}-20n^{3}\right)\psi_{m+n}-\frac{23\pi}{3k}i\Omega_{m+n}^{\left(11/2\right)}\nonumber \\
 &  & +\frac{\pi}{3k}\left(23m-82n\right)\Omega_{m+n}^{\left(9/2\right)}\,,\nonumber \\
i\left\{ \psi_{m},\psi_{n}\right\}  & = & 3\mathcal{W}_{m+n}+\frac{3}{2}\left(m^{2}-\frac{4}{3}mn+n^{2}\right)\mathcal{P}_{m+n}+\frac{9\pi}{k}\Omega_{m+n}^{\left(4\right)}+km^{4}\delta_{m+n,0}\,,\nonumber 
\end{eqnarray}
where $\Omega_{m}^{\left(l\right)}$, and $\Theta_{m}^{\left(l\right)}$
stand for the mode expansion of the following nonlinear terms:%
\footnote{The infinite-dimensional nonlinear algebras in eqs. \eqref{eq:InonuHyperBMS3}
and \eqref{eq:hyper-BMS-4}, correspond to different hypersymmetric
extensions of the BMS$_{3}$ algebra, being isomorphic to the Galilean
conformal algebra in two dimensions, and then relevant in the context
of non-relativistic holography \cite{Bagchi-Gopakumar}, \cite{Bagchi},
\cite{Bagchi.et.al-Flat Cosmology}, \cite{BDGS-cosmic}, \cite{BBGR-galilean}.%
}
\begin{eqnarray}
\Omega^{\left(4\right)} & = & \frac{1}{2}\mathcal{P}^{2}\,,\nonumber \\
\Theta^{\left(4\right)} & = & \mathcal{J}\mathcal{P}\,,\nonumber \\
\Omega^{\left(9/2\right)} & = & \frac{1}{2}\mathcal{P}\mathcal{\psi}\,,\nonumber \\
\Omega^{\left(11/2\right)} & = & \frac{27}{46}\mathcal{P}\mbox{\ensuremath{'}}\mathcal{\psi}\,,\\
\Omega^{\left(6\right)} & = & -\frac{7}{36}\mathcal{W}\mathcal{P}-\frac{2\pi}{3k}\mathcal{P}^{3}+\frac{295}{864}\left(\mathcal{P}\mbox{\ensuremath{'}}\right)^{2}+\frac{11}{27}\mathcal{P}\mathcal{P}\mbox{\ensuremath{''}}\,,\nonumber \\
\Theta^{\left(6\right)} & = & -\frac{7}{36}\left(\mathcal{V}\mathcal{P}+\mathcal{W}\mathcal{J}\right)-\frac{2\pi}{k}\mathcal{P}^{2}\mathcal{J}+\frac{295}{432}\mathcal{J}\mbox{\ensuremath{'}}\mathcal{P}\mbox{\ensuremath{'}}+\frac{11}{27}\left(\mathcal{J}\mathcal{P}\mbox{\ensuremath{''}}+\mathcal{P}\mathcal{J}\mbox{\ensuremath{''}}\right)+\frac{25}{72}i\psi\psi\mbox{\ensuremath{'}}\,.\nonumber 
\end{eqnarray}
Indeed, this asymptotic symmetry algebra is recovered from a contraction
that corresponds to a different flat limit of W$_{\left(2,4\right)}$$\oplus$W$_{\left(2,\frac{5}{2},4\right)}$,
as compared with the one in \ref{FlatLimit}. The flat limit is now
defined according to 
\[
\mathcal{P}_{n}=\frac{1}{\ell}\left(\mathcal{L}_{n}^{+}+\mathcal{L}_{-n}^{-}\right)\quad,\quad\mathcal{J}_{n}=\mathcal{L}_{n}^{+}-\mathcal{L}_{-n}^{-}\,,
\]
\begin{equation}
\mathcal{W}_{n}=\frac{1}{\ell}\left(\mathcal{U}_{n}^{+}+\mathcal{U}_{-n}^{-}\right)\quad,\quad\mathcal{V}_{n}=\mathcal{U}_{n}^{+}-\mathcal{U}_{-n}^{-}\quad,\quad\psi_{n}=\sqrt{\frac{6}{\ell}}\Psi_{n}^{+}\,,
\end{equation}
where the level also rescales as $\kappa=k\ell$. 

Moreover, once the modes $\mathcal{P}_{m}$ are shifted according
to \eqref{eq:Shift}, it is simple to verify that the wedge algebra
of \eqref{eq:hyper-BMS-4} reduces to the algebra of the gauge group
in \eqref{eq:hyper-Poincare-4}. 

From the anticommutator of the fermionic generators in \eqref{eq:hyper-BMS-4},
one then finds that the zero modes of the energy and the electric-like
spin-4 charge, $2\pi\mathcal{P}=\mathcal{P}_{0}$, $2\pi\mathcal{W}=\mathcal{W}_{0}$,
fulfill the following bounds 
\begin{equation}
3\mathcal{W}+\frac{9\pi}{2k}\mathcal{P}^{2}+5p^{2}\mathcal{P}+\frac{k}{2\pi}p^{4}\geq0\,,\label{eq:Bounds-spin4-1}
\end{equation}
which agree with the bounds in \cite{Hyper-AdS} in the flat limit.
It would then be interesting to explore different classes of solutions
endowed with electric-like spin-4 charge, including cosmological spacetimes
and solitonic-like configurations that fulfill the bounds \eqref{eq:Bounds-spin4-1},
as well as the hypersymmetric ones that should saturate them. Note
that since the bounds \eqref{eq:Bounds-spin4-1} factorize as
\begin{equation}
\left(p^{2}+\lambda_{\left[+\right]}^{2}\right)\left(p^{2}+\lambda_{\left[-\right]}^{2}\right)\geq0\,,\label{eq:Bounds-spin4}
\end{equation}
it is natural to expect that the eigenvalues of the holonomy of the
dynamical gauge field $a_{\phi}$ along an angular cycle, for the
class of solutions aforementioned, have to be given by
\begin{equation}
\lambda_{\left[\pm\right]}^{2}=\frac{5\pi}{k}\left(\mathcal{P}\pm\frac{4}{5}\sqrt{\mathcal{P}^{2}-\frac{3k}{8\pi}\mathcal{W}}\right)\,.
\end{equation}
In the case of solitonic-like solutions, these eigenvalues should
then correspond to a couple of purely imaginary integers, that become
related for the class of configurations that fulfill the bounds \eqref{eq:Bounds-spin4},
saturating just some of them.

As a final remark, since the hyper-Poincaré group actually exists
for $d\geq3$ spacetime dimensions \cite{FMT-3D-HYGRA}, it would
be interesting to explore whether similar results as the ones obtained
here could extend to higher-dimensional spacetimes. In this sense,
despite the no-go results in four dimensions \cite{AD-ConsistencyProb},
\cite{BVHVN}, \cite{AD2}, \cite{Porrati}, \cite{Bekaert-Boulanger-Sundell},
some interesting results have been recently found in the case of hypergravity
at the noninteracting level \cite{BHHL-hypergravity}. Whether these
results correspond to a suitable weak field limit of Vasiliev higher
spin gravity \cite{Vasiliev-higher spin}, \cite{Didenko-Skvortsov},
or another theory that has yet to be found, remains as an open question.

\acknowledgments We thank M. Henneaux, A. Pérez and D. Tempo for
helpful discussions. We also thank the International Solvay Institutes
and the ULB for warm hospitality. O.F. and R.T. wish to thank Amilcar
de Queiroz, Emanuele Orazi, and the organizers of the school and conference
``Theoretical Frontiers in Black Holes and Cosmology'', hosted by
IIP-UFRN, during June 2015, for the opportunity of presenting this
work. O.F. thanks Conicyt for financial support. This research has
been partially supported by Fondecyt grants Nº 1130658, 1121031, 3150448.
The Centro de Estudios Científicos (CECs) is funded by the Chilean
Government through the Centers of Excellence Base Financing Program
of Conicyt.

\appendix

\section{Conventions \label{A}}

We follow the conventions of \cite{FMT-3D-HYGRA}. The orientation
is chosen to be such that the Levi-Civita symbol fulfills $\varepsilon_{012}=1$,
and the Minkowski metric is assumed to be non-diagonal, whose only
nonvanishing components read $\eta_{01}=\eta_{10}=\eta_{22}=1$. Round
brackets correspond to symmetrization of the indices enclosed by them,
so that
\begin{equation}
X^{\left(a\right|}Y^{b}Z^{\left|c\right)}=X^{a}Y^{b}Z^{c}+X^{c}Y^{b}Z^{a}\,.
\end{equation}
Note that the three-dimensional Dirac matrices, that satisfy the Clifford
algebra $\left\{ \Gamma_{a},\Gamma_{b}\right\} =2\eta_{ab}$, fulfill
the following identity: 
\begin{equation}
\Gamma^{a}\Gamma^{b}\Gamma^{c}=\varepsilon^{abc}+\eta^{ab}\Gamma^{c}+\eta^{bc}\Gamma^{a}-\eta^{ac}\Gamma^{b}\,.
\end{equation}
The generators of the Lorentz group, in the spinorial and vector (adjoint)
representations, are assumed to be given by $\left(J_{a}\right)_{\beta}^{\alpha}=\frac{1}{2}\left(\Gamma_{a}\right)_{\beta}^{\alpha}$,
and $\left(J_{a}\right)_{c}^{b}=-\varepsilon_{a\,\,\,\, c}^{\,\,\, b}$,
respectively.

The three-dimensional $\Gamma$-matrices are chosen as
\begin{equation}
\Gamma_{0}=\frac{1}{\sqrt{2}}\left(\sigma_{1}+i\sigma_{2}\right)\quad,\quad\Gamma_{1}=\frac{1}{\sqrt{2}}\left(\sigma_{1}-i\sigma_{2}\right)\quad,\quad\Gamma_{2}=\sigma_{3}\,,
\end{equation}
 where $\sigma_{i}$ stand for the Pauli matrices:
\begin{equation}
\sigma_{1}=\left(\begin{array}{cc}
0 & 1\\
1 & 0
\end{array}\right)\quad,\quad\sigma_{2}=\left(\begin{array}{cc}
0 & -i\\
i & 0
\end{array}\right)\quad,\quad\sigma_{3}=\left(\begin{array}{cc}
1 & 0\\
0 & -1
\end{array}\right)\,.
\end{equation}
For a vector-spinor $\psi_{a}^{\alpha}$, with $\alpha=+,-$, and
$a=0,1,2$, the Majorana conjugate is defined as $\bar{\psi}_{\alpha a}=\psi_{a}^{\beta}C_{\beta\alpha}$,
where the charge conjugation matrix $C$, and its inverse are chosen
as
\begin{equation}
C_{\alpha\beta}=\left(\begin{array}{cc}
0 & -1\\
1 & 0
\end{array}\right)\quad,\quad C^{\alpha\beta}=\left(\begin{array}{cc}
0 & 1\\
-1 & 0
\end{array}\right)\,,
\end{equation}
so that $C^{T}=-C$, and $\left(C\Gamma_{a}\right)^{T}=C\Gamma_{a}$.

\section{Killing vector-spinors from an alternative approach \label{ExactKillingV-S}}

Killing vector-spinors $\epsilon_{a}^{\alpha}$ in \eqref{eq:Killing VS}
that are globally well defined can be recovered as a particular case
of the asymptotic symmetries discussed in section \ref{Asymptotics}.
Indeed, they are of the form $\epsilon_{a}^{\alpha}Q_{\alpha}^{a}=\lambda\left[0,0,\mathcal{E}\right]$,
with $\lambda$ given by \eqref{eq:lambda}. Hence, for the class
of bosonic configurations discussed in section \ref{Cosmology}, that
only carries the zero modes of the global charges, and their corresponding
chemical potentials are constant, the components of the Killing vector-spinors
can be written as
\begin{align}
\lambda\left[0,0,\mathcal{E}\right] & =\mathcal{E}\hat{\mathcal{Q}}_{\frac{3}{2}}-\mathcal{E}\mbox{\ensuremath{'}}\hat{\mathcal{Q}}_{\frac{1}{2}}-\frac{1}{2}\left(\frac{3\pi}{k}\mathcal{E}\mathcal{P}-\mathcal{E}\mbox{\ensuremath{''}}\right)\hat{\mathcal{Q}}_{-\frac{1}{2}}-\frac{\pi}{3k}\left(-\frac{7}{2}\mathcal{E}\mbox{\ensuremath{'}}\mathcal{P}+\frac{k}{2\pi}\mathcal{E}\mbox{\ensuremath{'''}}\right)\hat{\mathcal{Q}}_{-\frac{3}{2}}\,.\label{eq:lambdaE}
\end{align}
The requirements of invariance under hypersymmetry can then be read
from eqs. \eqref{eq:dfields}, \eqref{eq:ChiralityConditions}, so
that the Killing vector-spinor equation reduces to 
\begin{equation}
\delta\psi=-\frac{9\pi}{2k}\mathcal{P}^{2}\mathcal{E}+5\mathcal{P}\mathcal{E}\mbox{\ensuremath{''}}-\frac{k}{2\pi}\mathcal{E}\mbox{\ensuremath{''''}}=0\,,\label{eq:deltaPsi}
\end{equation}

\begin{equation}
\dot{\mathcal{E}}=\mu_{\mathcal{J}}\mathcal{E}\mbox{\ensuremath{'}}\,,\label{eq:epsilonPunto}
\end{equation}
which can be readily integrated. In fact, the solution of eq. \eqref{eq:deltaPsi}
is generically given by
\begin{equation}
\mathcal{E}=\mathcal{A}_{1}e^{\sqrt{\frac{\pi\mathcal{P}}{k}}\phi}+\mathcal{A}_{2}e^{-\sqrt{\frac{\pi\mathcal{P}}{k}}\phi}+\mathcal{A}_{3}e^{3\sqrt{\frac{\pi\mathcal{P}}{k}}\phi}+\mathcal{A}_{4}e^{-3\sqrt{\frac{\pi\mathcal{P}}{k}}\phi}\,,\label{eq:E}
\end{equation}
where $\mathcal{A}_{I}=\mathcal{A}_{I}\left(u\right)$ stand for four
arbitrary functions.

In the case of $\mathcal{P}>0$, $\mathcal{E}$ clearly cannot fulfill
neither antiperiodic nor periodic boundary conditions, and therefore,
cosmological spacetimes break all the hypersymmetries.

Note that if the energy vanishes $\left(\mathcal{P}=0\right)$, eq.
\eqref{eq:deltaPsi} integrates in a different way:
\begin{equation}
\mathcal{E}=\mathcal{A}_{0}+\mathcal{A}_{1}\phi+\mathcal{A}_{2}\phi^{2}+\mathcal{A}_{3}\phi^{3}\,.
\end{equation}
Periodicity then implies that $\mathcal{E}=\mathcal{A}_{0}\left(u\right)$,
while the remaining equation \eqref{eq:epsilonPunto} fixes the arbitrary
function to be a constant. Hence, vanishing energy configurations,
as the null orbifold, admit a single constant Killing vector-spinor.

Finally, for $\mathcal{P}=-kn^{2}/\pi<0$, eq. \eqref{eq:E} reads
\begin{equation}
\mathcal{E}=\mathcal{A}_{1}e^{in\phi}+\mathcal{A}_{1}^{*}e^{-in\phi}+\mathcal{A}_{3}e^{3in\phi}+\mathcal{A}_{3}^{*}e^{-3in\phi}\,,
\end{equation}
so that $n$ turns out to be a (half-)integer for fermions fulfilling
(anti)periodic boundary conditions. The remaining equation \eqref{eq:epsilonPunto}
then fixes the form of the arbitrary functions $\mathcal{A}_{I}\left(u\right)$,
and hence there are four independent Killing vector-spinors, determined
by
\begin{equation}
\mathcal{E}=\mathcal{E}_{1}e^{in\left(\mbox{\ensuremath{\mu}}_{\mathcal{J}}u+\phi\right)}+\mathcal{E}_{1}^{*}e^{-in\left(\mbox{\ensuremath{\mu}}_{\mathcal{J}}u+\phi\right)}+\mathcal{E}_{3}e^{3in\left(\mbox{\ensuremath{\mu}}_{\mathcal{J}}u+\phi\right)}+\mathcal{E}_{3}^{*}e^{-3in\left(\mbox{\ensuremath{\mu}}_{\mathcal{J}}u+\phi\right)}\,.
\end{equation}
 It is worth pointing out that Minkowski spacetime, which corresponds
to $n^{2}=1/4$, is the only one of this class that fulfills all the
hypersymmetry bounds \eqref{eq:boundsmn}, saturating precisely four
of them in the case of antiperiodic boundary conditions. Indeed, the
remaining solutions of this sort, in spite of possessing the maximum
number of hypersymmetries, become manifestly excluded by the bounds.
In the case of periodic boundary conditions, the null orbifold also
satisfies the bounds, but it saturates only one of them.

\section{Hyper-Poincaré algebra with fermionic generators of spin $n+\frac{1}{2}$\label{hyperPoincareSpinS}}

The extension of the Poincaré algebra in the generic case includes
fermionic tensor-spinor generators $Q_{\alpha}^{a_{1}\dots a_{n}}$,
being completely symmetric in the vector indices, and $\Gamma$-traceless,
i. e., $Q^{a_{1}\dots a_{n}}\Gamma_{a_{1}}=0$ \cite{FMT-3D-HYGRA}.
The Maurer-Cartan 1-form, 
\begin{equation}
\Omega=\rho^{a}P_{a}+\tau^{a}J_{a}+\chi_{a_{1}\dots a_{n}}^{\alpha}Q_{\alpha}^{a_{1}\dots a_{n}},\label{eq:MC-form}
\end{equation}
corresponds to a flat connection that fulfills
\begin{align}
d\tau^{a} & =-\frac{1}{2}\epsilon^{abc}\tau_{b}\tau_{c}\,,\label{eq:MC-algebra}\\
d\rho^{a} & =-\epsilon^{abc}\tau_{b}\rho_{c}+\frac{1}{2}\left(n+\frac{1}{2}\right)i\bar{\chi}_{a_{1}\dots a_{n}}\Gamma^{a}\chi^{a_{1}\dots a_{n}}\,,\nonumber \\
d\chi^{a_{1}\dots a_{n}} & =-\left(n+\frac{1}{2}\right)\tau^{b}\Gamma_{b}\chi^{a_{1}\dots a_{n}}+\tau_{b}\Gamma^{\left(a_{1}\right|}\chi^{\left|a_{2}\dots a_{n}\right)b}\,,\nonumber 
\end{align}
where $\chi_{a_{1}\dots a_{n}}$ is $\Gamma$-traceless, and completely
symmetric in the vector indices. Apart from $I_{1}=P^{a}P_{a}$, this
algebra admits an additional Casimir operator given by
\begin{equation}
I_{2}=2J^{a}P_{a}+Q_{\alpha a_{1}\dots a_{n}}C^{\alpha\beta}Q_{\beta}^{a_{1}\dots a_{n}}.\label{eq:Casimir-n}
\end{equation}

\section{Asymptotic hypersymmetry algebra \label{AppendixD}}

\subsection{Spin-$3/2$ fields (supergravity) \label{supergravity}}

In our conventions, the super-Poincaré algebra with $\mathcal{N}=1$
reads
\begin{eqnarray}
\left[J_{a},J_{b}\right] & = & \varepsilon_{abc}J^{c}\,,\nonumber \\
\left[J_{a},P_{b}\right] & = & \varepsilon_{abc}P^{c}\,,\\
\left[J_{a},Q_{\alpha}\right] & = & \frac{1}{2}\left(\Gamma_{a}\right)_{\,\,\,\,\alpha}^{\beta}Q_{\beta b}\,,\label{eq:superPoincare}\\
\left\{ Q_{\alpha},Q_{\beta}\right\}  & = & -\frac{1}{2}\left(C\Gamma^{c}\right)_{\alpha\beta}P_{c}\,,\nonumber 
\end{eqnarray}
so that changing the basis according to
\begin{gather}
\hat{\mathcal{J}}_{-1}=-2J_{0}\quad,\quad\hat{\mathcal{J}}_{1}=J_{1}\quad,\quad\hat{\mathcal{J}}_{0}=J_{2}\,,\nonumber \\
\hat{\mathcal{P}}_{-1}=-2P_{0}\quad,\quad\hat{\mathcal{P}}_{1}=P_{1}\quad,\quad\hat{\mathcal{P}}_{0}=P_{2}\,,\\
\hat{\mathcal{Q}}_{-\frac{1}{2}}=2^{\frac{3}{4}}Q_{+}\quad,\quad\hat{\mathcal{Q}}_{\frac{1}{2}}=2^{\frac{1}{4}}Q_{-}\,,\nonumber 
\end{gather}
it acquires the following form
\begin{eqnarray}
\left[\hat{\mathcal{J}}_{m},\hat{\mathcal{J}}_{n}\right] & = & \left(m-n\right)\hat{\mathcal{J}}_{m+n}\,,\nonumber \\
\left[\hat{\mathcal{J}}_{m},\hat{\mathcal{P}}_{n}\right] & = & \left(m-n\right)\hat{\mathcal{P}}_{m+n}\,,\nonumber \\
\left[\hat{\mathcal{J}}_{m},\hat{\mathcal{Q}}_{p}\right] & = & \left(\frac{m}{2}-p\right)\hat{\mathcal{Q}}_{m+p}\,,\label{eq:hyperPoincareModos-1-1}\\
\left\{ \hat{\mathcal{Q}}_{p},\hat{\mathcal{Q}}_{q}\right\}  & = & -\hat{\mathcal{P}}_{p+q}\,.\nonumber 
\end{eqnarray}
with $m,n=0,\pm1$, and $p,q=\pm\frac{1}{2}$.

The asymptotic form of the dynamical gauge field then reads
\begin{equation}
a_{\phi}=\hat{\mathcal{J}_{1}}-\frac{\pi}{k}\left(\mathcal{P}\hat{\mathcal{J}}_{-1}+\mathcal{J}\hat{\mathcal{P}}_{-1}+\psi\hat{\mathcal{Q}}_{-\frac{1}{2}}\right)\,,
\end{equation}
which is mapped into itself under gauge transformations generated
by
\begin{equation}
\lambda=T\hat{\mathcal{P}}_{1}+Y\hat{\mathcal{J}}_{1}+\mathcal{E}\hat{\mathcal{Q}}_{\frac{1}{2}}+\eta_{\left(\frac{1}{2}\right)}\left[T,Y,\mathcal{E}\right]\,,
\end{equation}
with
\begin{eqnarray}
\eta_{\left(\frac{1}{2}\right)}\left[T,Y,\mathcal{E}\right] & = & -T\mbox{\ensuremath{'}}\hat{\mathcal{P}}_{0}-Y\mbox{\ensuremath{'}}\hat{\mathcal{J}}_{0}-\left(\mathcal{E}\mbox{\ensuremath{'}}+\frac{\pi Y\psi}{k}\right)\hat{\mathcal{Q}}_{-\frac{1}{2}}+\frac{1}{2}\left(Y\mbox{\ensuremath{''}}-\frac{2\pi Y\mathcal{P}}{k}\right)\hat{\mathcal{J}}_{-1}\nonumber \\
 &  & +\frac{1}{2}\left(T\mbox{\ensuremath{''}}-\frac{2\pi T\mathcal{P}}{k}-\frac{2\pi Y\mathcal{J}}{k}+\frac{i\pi\mathcal{E}\psi}{k}\right)\hat{\mathcal{P}}_{-1}\,,
\end{eqnarray}
provided the fields transform according to:
\begin{eqnarray}
\delta\mathcal{P} & = & 2\mathcal{P}Y\mbox{\ensuremath{'}}+\mbox{\ensuremath{\mathcal{P}}}\mbox{\ensuremath{'}}Y-\frac{k}{2\pi}Y\mbox{\ensuremath{'''}}\,,\nonumber \\
\delta\mathcal{\mathcal{J}} & = & 2\mathcal{J}Y\mbox{\ensuremath{'}}+\mathcal{J}\mbox{\ensuremath{'}}Y+2\mathcal{P}T\mbox{\ensuremath{'}}+\mathcal{P}\mbox{\ensuremath{'}}T-\frac{k}{2\pi}T\mbox{\ensuremath{'''}}+\frac{3}{2}i\psi\mathcal{E}\mbox{\ensuremath{'}}+\frac{1}{2}i\psi\mbox{\ensuremath{'}}\mathcal{E}\,,\\
\delta\psi & = & \frac{3}{2}\psi Y\mbox{\ensuremath{'}}+\psi\mbox{\ensuremath{'}}Y-\mathcal{P}\mathcal{E}+\frac{k}{\pi}\mathcal{E}\mbox{\ensuremath{''}}\,.\nonumber 
\end{eqnarray}
Once expanded in modes, the asymptotic symmetry algebra is found to
be given by
\begin{eqnarray}
i\left\{ \mathcal{J}_{m},\mathcal{J}_{n}\right\}  & = & \left(m-n\right)\mathcal{J}_{m+n}\,,\nonumber \\
i\left\{ \mathcal{J}_{m},\mathcal{P}_{n}\right\}  & = & \left(m-n\right)\mathcal{P}_{m+n}+km^{3}\delta_{m+n,0}\,,\nonumber \\
i\left\{ \mathcal{J}_{m},\mathcal{\psi}_{n}\right\}  & = & \left(\frac{m}{2}-n\right)\mathcal{\psi}_{m+n}\,,\label{eq:superBMS}\\
i\left\{ \psi_{m},\mathcal{\psi}_{n}\right\}  & = & \mathcal{P}_{m+n}+2km^{2}\delta_{m+n,0}\,,\nonumber 
\end{eqnarray}
and hence in the case of fermions that fulfill periodic boundary conditions,
the energy $\mathcal{P}=\frac{\mathcal{P}_{0}}{2\pi}$ is bounded
to be nonnegative, 
\begin{equation}
\mathcal{P}\geq0\,,
\end{equation}
while in the case of fermions subject to antiperiodic boundary conditions,
the energy fulfills
\begin{equation}
\frac{1}{4}+\frac{\pi\mathcal{P}}{k}\geq0\,.
\end{equation}
These results agree with the ones in \cite{BDMT}.

\subsection{Spin-$7/2$ fields\label{Spin5/2}}

In the case of fermionic generators of (conformal) spin $\Delta=7/2$,
the hyper-Poincaré algebra can be written as
\begin{eqnarray}
\left[\hat{\mathcal{J}}_{m},\hat{\mathcal{J}}_{n}\right] & = & \left(m-n\right)\hat{\mathcal{J}}_{m+n}\,,\nonumber \\
\left[\hat{\mathcal{J}}_{m},\hat{\mathcal{P}}_{n}\right] & = & \left(m-n\right)\hat{\mathcal{P}}_{m+n}\,,\\
\left[\hat{\mathcal{J}}_{m},\hat{\mathcal{Q}}_{p}\right] & = & \left(\frac{5m}{2}-p\right)\hat{\mathcal{Q}}_{m+p}\,,\\
\left\{ \hat{\mathcal{Q}}_{p},\hat{\mathcal{Q}}_{q}\right\}  & = & f_{p,q}^{\left(5/2\right)}\hat{\mathcal{P}}_{p+q}\,,\nonumber 
\end{eqnarray}
with
\begin{eqnarray}
f_{p,q}^{\left(5/2\right)} & =- & \frac{1}{192}\left[80\left(p^{4}+q^{4}\right)-128\left(p^{3}q+pq^{3}\right)\right.\nonumber \\
 &  & \left.+144p^{2}q^{2}-620\left(p^{2}+q^{2}\right)+832pq+675\right]\,,
\end{eqnarray}
where $m,n=0,\pm1$, and $p,q=\pm\frac{1}{2},\pm\frac{3}{2},\pm\frac{5}{2}$.

The asymptotic behaviour of the dynamical gauge field now reads
\begin{equation}
a_{\phi}=\hat{\mathcal{J}}_{1}-\frac{\pi}{k}\left(\mathcal{J}\hat{\mathcal{P}}_{-1}+\mathcal{P}\hat{\mathcal{J}}_{-1}+\frac{\psi}{10}\hat{\mathcal{Q}}_{-\frac{5}{2}}\right)\,,
\end{equation}
so that the asymptotic symmetries are now spanned by
\begin{equation}
\lambda=T\hat{\mathcal{P}}_{1}+Y\hat{\mathcal{J}}_{1}+\mathcal{E}\hat{\mathcal{Q}}_{\frac{1}{2}}+\eta_{\left(\frac{5}{2}\right)}\left[T,Y,\mathcal{E}\right]\,,
\end{equation}
with
\begin{eqnarray}
\eta_{\left(\frac{5}{2}\right)}\left[T,Y,\mathcal{E}\right] & = & -T\mbox{\ensuremath{'}}\hat{\mathcal{P}}_{0}-Y\mbox{\ensuremath{'}}\hat{\mathcal{J}}_{0}-\mathcal{E}\mbox{\ensuremath{'}}\hat{\mathcal{Q}}_{\frac{3}{2}}+\frac{1}{2}\left(T\mbox{\ensuremath{''}}-\frac{2\pi T\mathcal{P}}{k}-\frac{2\pi Y\mathcal{J}}{k}-\frac{5i\pi\mathcal{E}\psi}{k}\right)\hat{\mathcal{P}}_{-1}\nonumber \\
 &  & +\frac{1}{2}\left(Y\mbox{\ensuremath{''}}-\frac{2\pi Y\mathcal{P}}{k}\right)\hat{\mathcal{J}}_{-1}+\frac{1}{2}\left(\mathcal{E}\mbox{\ensuremath{''}}-\frac{5\pi\mathcal{E}\mathcal{P}}{k}\right)\hat{\mathcal{Q}}_{\frac{1}{2}}\nonumber \\
 &  & -\frac{1}{6}\left(\mathcal{E}\mbox{\ensuremath{'''}}-\frac{13\pi\mathcal{E}\mbox{\ensuremath{'}}\mathcal{P}}{k}-\frac{5\pi\mathcal{E}\mathcal{P}\mbox{\ensuremath{'}}}{k}\right)\hat{\mathcal{Q}}_{-\frac{1}{2}}\\
 &  & +\frac{1}{24}\left(\mathcal{E}^{\left(4\right)}+\frac{45\pi^{2}\mathcal{E}\mathcal{P}{}^{2}}{k^{2}}-\frac{22\pi\mathcal{E}\mbox{\ensuremath{''}}\mathcal{P}}{k}-\frac{18\pi\mathcal{E}\mbox{\ensuremath{'}}\mathcal{P}\mbox{\ensuremath{'}}}{k}-\frac{5\pi\mathcal{E}\mathcal{P}\mbox{\ensuremath{''}}}{k}\right)\hat{\mathcal{Q}}_{-\frac{3}{2}}\nonumber \\
 &  & -\frac{1}{120}\left(\mathcal{E}^{(5)}+\frac{149\pi^{2}\mathcal{E}\mbox{\ensuremath{'}}\mathcal{P}^{2}}{k^{2}}+\frac{12\pi Y\psi}{k}-\frac{30\pi\mathcal{E}\mbox{\ensuremath{'''}}\mathcal{P}}{k}\right.\nonumber \\
 &  & \left.+\frac{130\pi^{2}\mathcal{E}\mathcal{P}\mathcal{P}\mbox{\ensuremath{'}}}{k^{2}}-\frac{40\pi\mathcal{E}\mbox{\ensuremath{''}}\mathcal{P}\mbox{\ensuremath{'}}}{k}-\frac{23\pi\mathcal{E}\mbox{\ensuremath{'}}\mathcal{P}\mbox{\ensuremath{''}}}{k}-\frac{5\pi\mathcal{E}\mathcal{P}\mbox{\ensuremath{'''}}}{k}\right)\hat{\mathcal{Q}}_{-\frac{5}{2}}\,,\nonumber 
\end{eqnarray}
and the transformation law of the fields is given by
\begin{eqnarray}
\delta\mathcal{P} & = & 2\mathcal{P}Y\mbox{\ensuremath{'}}+\mbox{\ensuremath{\mathcal{P}}}\mbox{\ensuremath{'}}Y-\frac{k}{2\pi}Y\mbox{\ensuremath{'''}}\,,\nonumber \\
\delta\mathcal{\mathcal{J}} & = & 2\mathcal{J}Y\mbox{\ensuremath{'}}+\mathcal{J}\mbox{\ensuremath{'}}Y+2\mathcal{P}T\mbox{\ensuremath{'}}+\mathcal{P}\mbox{\ensuremath{'}}T-\frac{k}{2\pi}T\mbox{\ensuremath{'''}}+\frac{7}{2}i\psi\mathcal{E}\mbox{\ensuremath{'}}+\frac{5}{2}i\psi\mbox{\ensuremath{'}}\mathcal{E}\,,\\
\delta\psi & = & \frac{7}{2}\psi Y\mbox{\ensuremath{'}}+\psi\mbox{\ensuremath{'}}Y+\left(-\frac{75\pi^{2}\mathcal{P}^{3}}{4k^{2}}+\frac{65\pi\left(\mathcal{P}\mbox{\ensuremath{'}}\right)^{2}}{6k}+\frac{155\pi\mathcal{P}\mathcal{P}\mbox{\ensuremath{''}}}{12k}-\frac{5}{12}\mathcal{P}^{(4)}\right)\mathcal{E}\nonumber \\
 &  & +\left(\frac{259\pi\mathcal{P}\mathcal{P}\mbox{\ensuremath{'}}}{6k}-\frac{7}{3}\mathcal{P}\mbox{\ensuremath{'''}}\right)\mathcal{E}\mbox{\ensuremath{'}}+\left(\frac{259\pi\mathcal{P}^{2}}{12k}-\frac{21}{4}\mathcal{P}\mbox{\ensuremath{''}}\right)\mathcal{E}\mbox{\ensuremath{''}}-\frac{35}{6}\mathcal{E}\mbox{\ensuremath{'''}}\mathcal{P}\mbox{\ensuremath{'}}-\frac{35}{12}\mathcal{E}^{(4)}\mathcal{P}+\frac{k\mathcal{E}^{(6)}}{12\pi}\,.\nonumber 
\end{eqnarray}
The Poisson brackets of the asymptotic symmetry algebra in this case
read
\begin{eqnarray}
i\left\{ \mathcal{J}_{m},\mathcal{J}_{n}\right\}  & = & \left(m-n\right)\mathcal{J}_{m+n}\,,\nonumber \\
i\left\{ \mathcal{J}_{m},\mathcal{P}_{n}\right\}  & = & \left(m-n\right)\mathcal{P}_{m+n}+km^{3}\delta_{m+n,0}\,,\nonumber \\
i\left\{ \mathcal{J}_{m},\mathcal{\psi}_{n}\right\}  & = & \left(\frac{5m}{2}-n\right)\mathcal{\psi}_{m+n}\,,\label{eq:hyperBMS7/2}\\
i\left\{ \psi_{m},\mathcal{\psi}_{n}\right\}  & = & \frac{1}{12}\left(5m^{4}-8m^{3}n+9m^{2}n^{2}-8mn^{3}+5n^{4}\right)\mathcal{P}_{m+n}\nonumber \\
 &  & +\frac{1}{48k}\left(155m^{2}-208mn+155n^{2}\right)\sum_{q}\mathcal{P}_{m+n-q}\mathcal{P}_{q}\nonumber \\
 &  & +\frac{75}{16k^{2}}\sum_{q}\mathcal{P}_{m+n-q-r}\mathcal{P}_{q}\mathcal{P}_{r}+\frac{k}{6}m^{6}\delta_{m+n,0}+\Xi_{m+n}^{\left(5/2\right)}\,,\nonumber 
\end{eqnarray}
where $\Xi_{m}^{\left(5/2\right)}=\int\Xi^{\left(5/2\right)}e^{-im\phi}d\phi$
stands for the mode expansion of
\begin{equation}
\Xi^{\left(5/2\right)}=\frac{25\pi\left(\mathcal{P}\mbox{\ensuremath{'}}\right)^{2}}{12k}\,.\label{eq:Xi52}
\end{equation}
The anticommutator of the fermionic charges then implies that the
energy has to fulfill the following bounds
\begin{equation}
\left(p^{2}+\frac{25\pi\mathcal{P}}{k}\right)\left(p^{2}+\frac{9\pi\mathcal{P}}{k}\right)\left(p^{2}+\frac{\pi\mathcal{P}}{k}\right)\geq0\,,
\end{equation}
with $p$ given by a (half-)integer for fermions that fulfill (anti)periodic
boundary conditions.

\subsection{Spin-$9/2$ fields\label{Spin7/2}}

The hyper-Poincaré algebra with fermionic generators of (conformal)
spin $\Delta=9/2$ is described by 
\begin{eqnarray}
\left[\hat{\mathcal{J}}_{m},\hat{\mathcal{J}}_{n}\right] & = & \left(m-n\right)\hat{\mathcal{J}}_{m+n}\,,\nonumber \\
\left[\hat{\mathcal{J}}_{m},\hat{\mathcal{P}}_{n}\right] & = & \left(m-n\right)\hat{\mathcal{P}}_{m+n}\,,\\
\left[\hat{\mathcal{J}}_{m},\hat{\mathcal{Q}}_{p}\right] & = & \left(\frac{7m}{2}-p\right)\hat{\mathcal{Q}}_{m+p}\,,\\
\left\{ \hat{\mathcal{Q}}_{p},\hat{\mathcal{Q}}_{q}\right\}  & = & f_{p,q}^{\left(7/2\right)}\hat{\mathcal{P}}_{p+q}\,,\nonumber 
\end{eqnarray}
where
\begin{eqnarray}
f_{p,q}^{\left(7/2\right)} & = & \frac{1}{2304}\left[112\left(p^{6}+q^{6}\right)-192\left(p^{5}q+pq^{5}\right)+240\left(p^{4}q^{2}+p^{2}q^{4}\right)-256p^{3}q^{3}\right.\nonumber \\
 &  & -2240\left(p^{4}+q^{4}\right)+3616\left(p^{3}q+pq^{3}\right)-4080p^{2}q^{2}\\
 &  & \left.+11578\left(p^{2}+q^{2}\right)-15592pq-11025\right]\,,\nonumber 
\end{eqnarray}
and with $m,n=0,\pm1$, and $p,q=\pm\frac{1}{2},\pm\frac{3}{2},\pm\frac{5}{2},\pm\frac{7}{2}$.
\newpage

The asymptotic fall-off of the dynamical gauge field is now given
by
\begin{equation}
a_{\phi}=\hat{\mathcal{J}}_{1}-\frac{\pi}{k}\left(\mathcal{J}\hat{\mathcal{P}}_{-1}+\mathcal{P}\hat{\mathcal{J}}_{-1}-\frac{\psi}{35}\hat{\mathcal{Q}}_{-\frac{7}{2}}\right)\,,
\end{equation}
and the asymptotic symmetries turn out to be parametrized according
to
\begin{equation}
\lambda=T\hat{\mathcal{P}}_{1}+Y\hat{\mathcal{J}}_{1}+\mathcal{E}\hat{\mathcal{Q}}_{\frac{1}{2}}+\eta_{\left(\frac{7}{2}\right)}\left[T,Y,\mathcal{E}\right]\,,
\end{equation}
with
\begin{eqnarray}
\eta_{\left(\frac{7}{2}\right)}\left[T,Y,\mathcal{E}\right] & = & -T\mbox{\ensuremath{'}}\hat{\mathcal{P}}_{0}-Y\mbox{\ensuremath{'}}\hat{\mathcal{J}}_{0}-\mathcal{E}\mbox{\ensuremath{'}}\hat{\mathcal{Q}}_{\frac{5}{2}}+\frac{1}{2}\left(T\mbox{\ensuremath{''}}-\frac{2\pi T\mathcal{P}}{k}-\frac{2\pi Y\mathcal{J}}{k}-\frac{7i\pi\mathcal{E}\psi}{k}\right)\hat{\mathcal{P}}_{-1}\nonumber \\
 &  & +\frac{1}{2}\left(Y\mbox{\ensuremath{''}}-\frac{2\pi Y\mathcal{P}}{k}\right)\hat{\mathcal{J}}_{-1}+\frac{1}{2}\left(\mathcal{E}\mbox{\ensuremath{''}}-\frac{7\pi\mathcal{E}\mathcal{P}}{k}\right)\hat{\mathcal{Q}}_{\frac{3}{2}}-\frac{1}{6}\left(\mathcal{E}\mbox{\ensuremath{'''}}-\frac{19\pi\mathcal{E}\mbox{\ensuremath{'}}\mathcal{P}}{k}-\frac{7\pi\mathcal{E}\mathcal{P}\mbox{\ensuremath{'}}}{k}\right)\hat{\mathcal{Q}}_{\frac{1}{2}}\nonumber \\
 &  & +\frac{1}{24}\left(\mathcal{E}^{(4)}+\frac{105\pi^{2}\mathcal{E}\mathcal{P}{}^{2}}{k^{2}}-\frac{34\pi\mathcal{E}\mbox{\ensuremath{''}}\mathcal{P}}{k}-\frac{26\pi\mathcal{E}\mbox{\ensuremath{'}}\mathcal{P}\mbox{\ensuremath{'}}}{k}-\frac{7\pi\mathcal{E}\mathcal{P}\mbox{\ensuremath{''}}}{k}\right)\hat{\mathcal{Q}}_{-\frac{1}{2}}\nonumber \\
 &  & -\frac{1}{120}\left(\mathcal{E}^{(5)}+\frac{409\pi^{2}\mathcal{E}\mbox{\ensuremath{'}}\mathcal{P}{}^{2}}{k^{2}}+\frac{322\pi^{2}\mathcal{E}\mathcal{P}\mathcal{P}\mbox{\ensuremath{'}}}{k^{2}}-\frac{50\pi\mathcal{E}\mbox{\ensuremath{'''}}\mathcal{P}}{k}-\frac{60\pi\mathcal{E}\mbox{\ensuremath{''}}\mathcal{P}\mbox{\ensuremath{'}}}{k}\right.\nonumber \\
 &  & \left.-\frac{33\pi\mathcal{E}\mbox{\ensuremath{'}}\mathcal{P}\mbox{\ensuremath{''}}}{k}-\frac{7\pi\mathcal{E}\mathcal{P}\mbox{\ensuremath{'''}}}{k}\right)\hat{\mathcal{Q}}_{-\frac{3}{2}}+\frac{1}{720}\left(\mathcal{E}^{(6)}-\frac{1575\pi^{3}\mathcal{E}\mathcal{P}{}^{3}}{k^{3}}\right.\\
 &  & +\frac{919\pi^{2}\mathcal{E}\mbox{\ensuremath{''}}\mathcal{P}{}^{2}}{k^{2}}+\frac{1530\pi^{2}\mathcal{E}\mbox{\ensuremath{'}}\mathcal{P}\mathcal{P}\mbox{\ensuremath{'}}}{k^{2}}+\frac{427\pi^{2}\mathcal{E}\mathcal{P}\mathcal{P}\mbox{\ensuremath{''}}}{k^{2}}+\frac{322\pi^{2}\mathcal{E}\left(\mathcal{P}\mbox{\ensuremath{'}}\right)^{2}}{k^{2}}\nonumber \\
 &  & \left.-\frac{65\pi\mathcal{E}^{(4)}\mathcal{P}}{k}-\frac{110\pi\mathcal{E}^{(3)}\mathcal{P}\mbox{\ensuremath{'}}}{k}-\frac{93\pi\mathcal{E}\mbox{\ensuremath{''}}\mathcal{P}\mbox{\ensuremath{''}}}{k}-\frac{40\pi\mathcal{E}\mbox{\ensuremath{'}}\mathcal{P}\mbox{\ensuremath{'''}}}{k}-\frac{7\pi\mathcal{E}\mathcal{P}^{(4)}}{k}\right)\hat{\mathcal{Q}}_{-\frac{5}{2}}\nonumber \\
 &  & -\frac{1}{5040}\left(\mathcal{E}^{(7)}-\frac{6483\pi^{3}\mathcal{E}'\mbox{\ensuremath{'}}\mathcal{P}{}^{3}}{k^{3}}-\frac{8589\pi^{3}\mathcal{E}\mathcal{P}{}^{2}\mathcal{P}\mbox{\ensuremath{'}}}{k^{3}}+\frac{1519\pi^{2}\mathcal{E}\mbox{\ensuremath{'''}}\mathcal{P}{}^{2}}{k^{2}}\right.\nonumber \\
 &  & +\frac{4088\pi^{2}\mathcal{E}\mbox{\ensuremath{''}}\mathcal{P}\mathcal{P}\mbox{\ensuremath{'}}}{k^{2}}+\frac{2353\pi^{2}\mathcal{E}\mbox{\ensuremath{'}}\mathcal{P}\mathcal{P}\mbox{\ensuremath{''}}}{k^{2}}+\frac{1852\pi^{2}\mathcal{E}\mbox{\ensuremath{'}}\left(\mathcal{P}\mbox{\ensuremath{'}}\right)^{2}}{k^{2}}\nonumber \\
 &  & +\frac{511\pi^{2}\mathcal{E}\mathcal{P}\mathcal{P}\mbox{\ensuremath{'''}}}{k^{2}}+\frac{1071\pi^{2}\mathcal{E}\mathcal{P}\mbox{\ensuremath{'}}\mathcal{P}\mbox{\ensuremath{''}}}{k^{2}}-\frac{144\pi Y\psi}{k}-\frac{77\pi\mathcal{E}^{(5)}\mathcal{P}}{k}-\frac{175\pi\mathcal{E}^{(4)}\mathcal{P}\mbox{\ensuremath{'}}}{k}\nonumber \\
 &  & \left.-\frac{203\pi\mathcal{E}\mbox{\ensuremath{'''}}\mathcal{P}\mbox{\ensuremath{''}}}{k}-\frac{133\pi\mathcal{E}\mbox{\ensuremath{''}}\mathcal{P}\mbox{\ensuremath{'''}}}{k}-\frac{47\pi\mathcal{E}\mbox{\ensuremath{'}}\mathcal{P}^{(4)}}{k}-\frac{7\pi\mathcal{E}\mathcal{P}^{(5)}}{k}\right)\hat{\mathcal{Q}}_{-\frac{7}{2}}\,,\nonumber 
\end{eqnarray}
so that the fields transform as
\begin{eqnarray}
\delta\mathcal{P} & = & 2\mathcal{P}Y\mbox{\ensuremath{'}}+\mbox{\ensuremath{\mathcal{P}}}\mbox{\ensuremath{'}}Y-\frac{k}{2\pi}Y\mbox{\ensuremath{'''}}\,,\nonumber \\
\delta\mathcal{\mathcal{J}} & = & 2\mathcal{J}Y\mbox{\ensuremath{'}}+\mathcal{J}\mbox{\ensuremath{'}}Y+2\mathcal{P}T\mbox{\ensuremath{'}}+\mathcal{P}\mbox{\ensuremath{'}}T-\frac{k}{2\pi}T\mbox{\ensuremath{'''}}+\frac{9}{2}i\psi\mathcal{E}\mbox{\ensuremath{'}}+\frac{7}{2}i\psi\mbox{\ensuremath{'}}\mathcal{E}\,,
\end{eqnarray}
\begin{eqnarray}
\delta\psi & = & \frac{9}{2}\psi Y\mbox{\ensuremath{'}}+\psi\mbox{\ensuremath{'}}Y+\left(-\frac{1225\pi^{3}\mathcal{P}^{4}}{16k^{3}}+\frac{5789\pi^{2}\mathcal{P}\mbox{\ensuremath{''}}\mathcal{P}^{2}}{72k^{2}}+\frac{2429\pi^{2}\left(\mathcal{P}\mbox{\ensuremath{'}}\right)^{2}\mathcal{P}}{18k^{2}}\right.\nonumber \\
 &  & \left.-\frac{35\pi\mathcal{P}^{(4)}\mathcal{P}}{9k}-\frac{119\pi\left(\mathcal{P}\mbox{\ensuremath{''}}\right)^{2}}{16k}-\frac{791\pi\mathcal{P}\mbox{\ensuremath{'}}\mathcal{P}\mbox{\ensuremath{'''}}}{72k}+\frac{7}{144}\mathcal{P}^{(6)}\right)\mathcal{E}+\nonumber \\
 &  & +\left(\frac{3229\pi^{2}\mathcal{P}\mbox{\ensuremath{'}}\mathcal{P}{}^{2}}{12k^{2}}-\frac{131\pi\mathcal{P}\mbox{\ensuremath{'''}}\mathcal{P}}{6k}-\frac{99\pi\mathcal{P}\mbox{\ensuremath{'}}\mathcal{P}\mbox{\ensuremath{''}}}{2k}+\frac{3}{8}\mathcal{P}^{(5)}\right)\mathcal{E}\mbox{\ensuremath{'}}\nonumber \\
 &  & +\left(\frac{3229\pi^{2}\mathcal{P}{}^{3}}{36k^{2}}-\frac{197\pi\mathcal{P}\mbox{\ensuremath{''}}\mathcal{P}}{4k}-\frac{165\pi\left(\mathcal{P}\mbox{\ensuremath{'}}\right)^{2}}{4k}+\frac{5}{4}\mathcal{P}^{(4)}\right)\mathcal{E}\mbox{\ensuremath{''}}\nonumber \\
 &  & +\left(\frac{7}{3}\mathcal{P}\mbox{\ensuremath{'''}}-\frac{329\pi\mathcal{P}\mathcal{P}\mbox{\ensuremath{'}}}{6k}\right)\mathcal{E}^{(3)}+\left(\frac{21}{8}\mathcal{P}\mbox{\ensuremath{''}}-\frac{329\pi\mathcal{P}{}^{2}}{24k}\right)\mathcal{E}^{(4)}\nonumber \\
 &  & +\frac{7}{4}\mathcal{E}^{(5)}\mathcal{P}\mbox{\ensuremath{'}}+\frac{7}{12}\mathcal{E}^{(6)}\mathcal{P}-\frac{k\mathcal{E}^{(8)}}{144\pi}\,.\nonumber 
\end{eqnarray}
The mode expansion of the asymptotic symmetry algebra is then given
by
\begin{eqnarray}
i\left\{ \mathcal{J}_{m},\mathcal{J}_{n}\right\}  & = & \left(m-n\right)\mathcal{J}_{m+n}\,,\nonumber \\
i\left\{ \mathcal{J}_{m},\mathcal{P}_{n}\right\}  & = & \left(m-n\right)\mathcal{P}_{m+n}+km^{3}\delta_{m+n,0}\,,\nonumber \\
i\left\{ \mathcal{J}_{m},\psi_{n}\right\}  & = & \left(\frac{7m}{2}-n\right)\mathcal{\psi}_{m+n}\,,\\
i\left\{ \psi_{m},\psi_{n}\right\}  & = & \frac{1}{144}\left(7m^{6}-12m^{5}n+15m^{4}n^{2}-16m^{3}n^{3}+15m^{2}n^{4}-12mn^{5}+7n^{6}\right)\mathcal{P}_{m+n}\nonumber \\
 &  & +\frac{1}{144k}\left(140m^{4}-226m^{3}n+255m^{2}n^{2}-226mn^{3}+140n^{4}\right)\sum_{q}\mathcal{P}_{m+n-q}\mathcal{P}_{q}\nonumber \\
 &  & +\frac{1}{864k^{2}}\left(5789m^{2}-7796mn+5789n^{2}\right)\sum_{q,r}\mathcal{P}_{m+n-q-r}\mathcal{P}_{q}\mathcal{P}_{r}\nonumber \\
 &  & +\frac{1225}{128k^{3}}\sum_{q,r,t}\mathcal{P}_{m+n-q-r}\mathcal{P}_{q}\mathcal{P}_{r}\mathcal{P}_{t}+\frac{k}{72}m^{8}\delta_{m+n,0}+\Xi_{m+n}^{\left(7/2\right)}\,,\nonumber 
\end{eqnarray}
where
\begin{equation}
\Xi_{m+n}^{\left(7/2\right)}=7\Theta_{m+n}+\left(329m^{2}-494mn+329n^{2}\right)\chi_{m+n}\,,\label{eq:Xi7/2}
\end{equation}
and $\Theta_{m}$ and $\chi_{m}$ correspond to the mode expansion
of 
\begin{equation}
\Theta=\frac{1596\pi^{2}\mathcal{P}\left(\mathcal{P}\mbox{\ensuremath{'}}\right)^{2}}{432k^{2}}+\frac{661\pi\left(\mathcal{P}\mbox{\ensuremath{''}}\right)^{2}}{432k}\quad,\quad\chi=\frac{\pi\left(\mathcal{P}\mbox{\ensuremath{'}}\right)^{2}}{144k}\,,
\end{equation}
respectively. 

The energy is then found to fulfill the following bounds
\begin{equation}
\left(p^{2}+\frac{49\pi\mathcal{P}}{k}\right)\left(p^{2}+\frac{25\pi\mathcal{P}}{k}\right)\left(p^{2}+\frac{9\pi\mathcal{P}}{k}\right)\left(p^{2}+\frac{\pi\mathcal{P}}{k}\right)\geq0\,,
\end{equation}
where according to the (anti)periodicity conditions for the fermions,
$p$ corresponds to a (half-)integer.

\end{document}